\documentclass[conference]{IEEEtran}
\IEEEoverridecommandlockouts
\usepackage{cite}
\usepackage{textcomp}
\def\BibTeX{{\rm B\kern-.05em{\sc i\kern-.025em b}\kern-.08em
    T\kern-.1667em\lower.7ex\hbox{E}\kern-.125emX}}

\usepackage{tikz}
\usepackage[utf8]{inputenc}

\usepackage{amssymb}
\usepackage{amsthm}
\usepackage{mathtools}
\usepackage{soul}
\usepackage{algorithm}
\usepackage[noend]{algpseudocode}
\usepackage{hyperref}
\let\labelindent\relax
\usepackage{enumitem}
\usepackage{subcaption}
\usepackage{indentfirst}
\usepackage{booktabs}
\usepackage{float}
\usepackage{todonotes}
\usepackage{xspace}
\usepackage{multirow}
\usepackage{tcolorbox}
\usepackage{enumitem}
\usepackage{cleveref}
\usepackage{pdfpages}

\input{macro}

\makeatletter
\newcommand{\linebreakand}{%
  \end{@IEEEauthorhalign}
  \hfill\mbox{}\par
  \mbox{}\hfill\begin{@IEEEauthorhalign}
}
\makeatother

\begin{document}

\title{Backdoor Detection through\\ Replicated Execution of Outsourced Training}

\author{\IEEEauthorblockN{Hengrui Jia\text{*}}
\IEEEauthorblockA{
\textit{University of Toronto, Vector Institute}\\
nickhengrui.jia@mail.utoronto.ca}
\and
\IEEEauthorblockN{Sierra Wyllie\text{*}}
\IEEEauthorblockA{
\textit{University of Toronto, Vector Institute}\\
sierra.wyllie@mail.utoronto.ca}
\and
\IEEEauthorblockN{Akram Bin Sediq}
\IEEEauthorblockA{
\textit{Ericsson Canada}\\
akram.bin.sediq@ericsson.com}
\linebreakand 
\IEEEauthorblockN{Ahmed Ibrahim}
\IEEEauthorblockA{
\textit{Ericsson Canada}\\
ahmed.a.ibrahim@ericsson.com}
\and
\IEEEauthorblockN{Nicolas Papernot}
\IEEEauthorblockA{
\textit{University of Toronto, Vector Institute}\\
nicolas.papernot@utoronto.ca}
}

\maketitle

\begin{abstract}
\looseness=-1
It is common practice to outsource the training of machine learning models to cloud providers. Clients who do so gain from the cloud's economies of scale, but implicitly assume trust: the server should not deviate from the client's training procedure. A malicious server may, for instance, seek to insert backdoors in the model. Detecting a backdoored model without prior knowledge of both the backdoor attack and its accompanying trigger remains a challenging problem. In this paper, we show that a client with access to multiple cloud providers can replicate a subset of training steps across multiple servers to detect deviation from the training procedure in a similar manner to differential testing. Assuming some cloud-provided servers are benign, we identify malicious servers by the substantial difference between model updates required for backdooring and those resulting from clean training. Perhaps the strongest advantage of our approach is its suitability to clients that have limited-to-no local compute capability to perform training; we leverage the existence of multiple cloud providers to identify malicious updates without expensive human labeling or heavy computation. We demonstrate the capabilities of our approach on an outsourced supervised learning task where $50\%$ of the cloud providers insert their own backdoor; our approach is able to correctly identify $99.6\%$ of them. In essence, our approach is successful because it replaces the signature-based paradigm taken by existing approaches with an anomaly-based detection paradigm. Furthermore, our approach is robust to several attacks from adaptive adversaries utilizing knowledge of our detection scheme. 
\blfootnote{\text{*}Co-first author.}
\end{abstract}

\section{Introduction}
\label{sec:intro}

Given the large computational footprint associated with machine learning, it is common to outsource the training of deep neural networks (DNNs) to cloud computing providers~\cite{mlaas2021}. Typically, the client uploads a dataset to the server, which then runs a mutually-agreed upon training algorithm on this data and returns a model to the client. This may expose the client to unfaithful executions of the training procedure by an untrustworthy server. For instance, the server may adversarially manipulate training updates to output a model containing a backdoor. 
While outsourced training was one of the earliest identified settings where models could be backdoored~\cite{guBadnets}, few defenses in the literature address this setting. Instead, much of the current literature focuses on defending against \textit{dataset-level adversaries} with the ability to modify the training dataset rather than the harder setting of \textit{training-level adversaries} who can modify the training process.

Previous defenses that attempt to mitigate dataset-level adversaries~\cite{wu2022backdoorbench,li2022backdoorsurvey} are inapplicable when model training is outsourced 
{as the trainers are the potential adversaries in our threat model, and they may not execute the training algorithms faithfully.}
It is also impossible to prevent backdoor insertion via training algorithms that provide certified robustness guarantees (such as~\cite{jia2021intrinsic,jia2022certified}) as the trainer may arbitrarily deviate from the training algorithm. Finally, methods that detect backdoors in the model after training often assume some properties based on existing attacks. In \Cref{ssec:existing_defenses} we show that these defenses can easily be evaded, \eg by a training-level adversary that lowers the learning rate. %

\looseness=-1
While outsourced training broadens the range and severity of attacks, it also provides a unique opportunity for backdoor attack detection through the multiplicity of available cloud providers. 
We demonstrate the feasibility of detecting backdoors by comparing the models returned by these different cloud providers as they execute an agreed-upon training algorithm. 
Our novel backdoor defense is called \name, or Replicate Training To Detect.
In our method, the client decomposes the training run into multiple smaller {\em sub-runs} and randomly replicates some of the sub-runs' execution across multiple servers owned by different cloud providers. Assuming non-colluding cloud providers, our key intuition is that benign servers will return similar updates, %
whereas adversaries will return updates that are out-of-distribution compared to benign updates. This is because adversaries aim to learn two tasks, the primary task, which benign servers also train for, and a unique backdoor task. A client can therefore detect a malicious server by comparing the server's returned model updates to the model updates returned by other servers during replicated training.

\looseness=-1
\name is analogous to differential testing~\cite{mckeeman1998differential}, a software testing technique that identifies bugs by comparing the outputs of multiple, similar programs given identical inputs.
However, comparing outputs of outsourced training is non-trivial due to the
stochasticity in model training (\eg caused by data augmentation, data ordering\footnote{In the limit, it is possible to perform backdoor attacks by only re-ordering the training datapoints~\cite{shumailov2021manipulating}.}, and noise in the hardware and compilers~\cite{hooker2022Randomness}) as it can cause even benign model updates to be dissimilar. 
We address this issue by (a) controlling the lengths of the aforementioned {\em sub-runs} to bound the randomness during training, and (b) analyzing various techniques for modeling the distributions of the model updates. For (b), we show that comparing model updates returned by different servers using model distance metrics, such as Zest~\cite{jia2022a}, significantly outperforms the approach taken by existing backdoor defenses. %
Indeed, our proposed solution follows an anomaly-based detection scheme (\ie any deviation from normal behavior is flagged as anomalous) whereas prior defenses are signature-based (\ie they make assumptions about what a backdoor should ``look like''). 

In \Cref{sec:evaluation}, we find that our method successfully detects malicious servers using several canonical attacks from the literature in both the computer vision and language domains. Namely, we study three variants of the BadNets ~\cite{guBadnets} attack, two variants of Wasserstein~\cite{khoaWasserstein} attack, as well as the WaNet~\cite{nguyenWaNet}, Bpp~\cite{wangBppAttack2022}, and Adaptive Blend~\cite{qi2023revisiting} attacks in computer vision. In language, we show \name's effectiveness against BadNets~\cite{guBadnets} and the Hidden Killer~\cite{qi2021hidden} attacks. Additionally, we find that our detector holds against the several adaptive adversaries we study; these adversaries know the entirety of the detector's mechanisms.
In summary, our contributions are the following:
\vspace{-1mm}
\begin{itemize}[leftmargin=*,noitemsep,topsep=0pt]
\item We analyze and explore an insufficiently studied yet practical threat model for backdoor attacks:  training outsourced to the server of an untrusted cloud provider. We show that existing state-of-the-art defenses are easily bypassed by a malicious server that that merely lowers the learning rate when training on backdoored data in \Cref{ssec:existing_defenses}.
\item \looseness=-1 We introduce a novel framework called \name, which leverages replicated executions of training sub-runs to detect backdoors inserted by malicious servers. It adopts an anomaly-based detection approach to detect malicious servers that deviate from the agreed-upon training dataset and algorithm -- which does not require making assumptions about the adversary's
{backdoor construction strategy.}
\looseness=-1
\item Empirical validation shows that on an outsourced supervised learning task where $50\%$ of the cloud providers are malicious and each inserts a different backdoor, \name is able to identify $99.6\%$ of the malicious cloud providers correctly.
\item We perform a comprehensive ablation study in \Cref{ssec:eval-ablation} showing that \name is robust to changes in the length of the sub-runs, the learning rate, and robust at different stages of training. We further show that \name is accurate even when training is outsourced to as few as three servers. 
\end{itemize}
\looseness=-1
A significant advantage of our approach is its suitability to clients with limited computational capability to perform model training or backdoor detection, as all training is outsourced to cloud providers. While outsourced and replicated training may incur a monetary cost, this may be traded-off with detection accuracy by replicating only a subset of the sub-runs or replicating them onto fewer servers. This makes our assumption of having access to multiple training providers more realistic. Rather than requiring separate detection and training mechanisms, our approach also yields the model the client initially wanted to train as a by-product of detection, further increasing efficiency. There are some defenses using cryptographic guarantees that force the server to execute a specific training protocol on specific data \eg within a secure enclave or by using verified computing~\cite{Walfish2013VerifyingCW}. Compared to our method, these cryptographic techniques are currently prohibitively computationally expensive, which limits their applicability. Instead, we show that our approach can scale to deep learning.

\section{Background and Threat Model}
\label{sec:background}

\subsection{Backdoor Attacks in Machine Learning}
\label{ssec:background_backdoor}
Backdoor attacks against machine learning models change the model's behaviour on backdoored data, which are data points that contain an adversary-chosen trigger. In conventional backdoor attacks, this is achieved by adding the triggers to a selection of points in the training set and changing their labels to encourage the models to associate the presence of the trigger with the backdoored label~\cite{guBadnets,chen2017targeted,shafahi2018poison,jagielski2018manipulating,bagdasaryan2020backdoor,wang2020attack}. 

A backdoor adversary can have varying capabilities: a weaker adversary might only be able to poison the dataset with backdoored data, while a stronger adversary can also arbitrarily alter the training algorithm itself. There are also backdoors that can be embedded in model architectures~\cite{architecturalbackdoors} or in the software~\cite{clifford2024impnet} and hardware stacks, which we deem out-of-scope as they are not data-based. In this work, we focus on the training level adversary. 
Next, we describe a few canonical backdoor attacks for supervised learning: 

\looseness=-1
\noindent \textbf{BadNets~\cite{guBadnets}.} The first backdoor attack to be proposed was BadNets~\cite{guBadnets}. It alters a few pixels in the input image to form the trigger. While the BadNets triggers are simple for the adversary to inject, they are also easily detected by both humans and algorithms by inspecting the input and latent spaces of the model~\cite{chenActivation}~\cite{TranSpectral}.  We consider three BadNets triggers: a small white square in the corner of the image, a white stripe, and an RGB flag (see~\Cref{fig:trigger} in Appendix~\ref{app:additional_figures}). 

\looseness=-1
\noindent \textbf{WaNet~\cite{nguyen2021wanet}.} WaNets is a backdoor attack that addresses a limitation of BadNets by making human-imperceptible triggers. This attack subtly warps the entire image according to an image-warping field that stretches and distorts the image. This trigger is more challenging for humans to detect but sufficient to achieve the desired backdoor effect once added to the input.

\looseness=-1
\noindent \textbf{BppAttack~\cite{wangBppAttack2022}.} The Bpp attack seeks to make stealthy triggers by lowering the bits-per-pixel needed to backdoor. The trigger is formed by projecting the image into a smaller color pallet and applying noise to reduce the obviousness of this quantization. Because the trigger is input-dependent, the attack uses contrastive adversarial training to ensure that the model learns to differentiate the triggers from other perturbations that can cause misprediction. Note that Bpp attackers need control over both the dataset and training process, making this attack particularly relevant to our threat model. 

\looseness=-1
\noindent \textbf{Adaptive Blend~\cite{qi2023revisiting}.} The Adaptive Blend attack designs triggers that are neither perceptible in the input space or in the model latent space (\ie the representations the model outputs after each hidden layer). To do so, the adversary need make two changes to the training dataset. 
First, rather than changing the labels of all training points that are perturbed to include a trigger, some of the triggered points retain their original label.
Second, the trigger is randomly masked out for each training point. 

\looseness=-1
\noindent\textbf{Wasserstein Backdoor~\cite{khoaWasserstein}.} 
The Wasserstein backdoor attack uses an ML model to output triggers rather than rely on manual trigger engineering. It uses a U-Net or autoencoder architecture where first half of the architecture is contractive and the second half gradually upsamples the representation back to the input dimensions. This model outputs the trigger to be added to an input, and is trained by minimizing the Wasserstein distance between the natural and backdoored data in the victim model's latent space~\cite{khoaWasserstein}. This requires an adversary with training-level access to the victim model, making it a strong and viable attack in our threat model.

\subsection{Backdoor Defenses in Machine Learning}
\label{ssec:background_detection}

\looseness=-1
Here, we overview three categories of backdoor defenses which we differentiate based on their level of data access and their assumptions. We find that only one of these categories is applicable to our threat model, and overview three of it's state-of-the defenses which we evaluate against later in this work.

The first category is blind backdoor removal techniques, which minimize the effect of any backdoors that may be present~\cite{gaoBackdoorReview}. These techniques include neuron pruning~\cite{liuFinePruning}, ignoring image style information~\cite{VasquezConFoc},  %
and smoothing the training set with noise~\cite{weberRAB}. Such methods often decrease model performance and rely on knowledge of the backdoor attack. To avoid these shortcomings, our method instead focuses on determining if a backdoor was inserted. 

The second category of defenses rely on the differences between clean and backdoored data or between the latent space representations of clean and backdoored models. For example, Spectral Signature~\cite{TranSpectral} and Activation Clustering~\cite{chenActivation} assume that backdoored data representations map to a separate cluster or are outliers (all with the same label) when compared to the clean data representations. 
However, these defenses can be bypassed by attacks like Adaptive Blend~\cite{qi2023revisiting} and Wasserstein Backdoor~\cite{khoaWasserstein} which optimize for stealthiness. They also require access to the backdoored dataset used to train  the model. However, this data is unavailable in our threat model because the client cannot trust that a server trained on the dataset that they claim to have trained on.

\looseness=-1
The final category of defenses only requires access to the model and a few correctly labeled samples, making them viable candidates for backdoor detection in our threat model. To identify triggers, they make some assumptions about the characteristics of the attacks or triggers, whereas we do not make such assumptions. We overview three defenses from this category:

\looseness=-1
\noindent\textbf{Neural Cleanse~\cite{Wang2019Neural}.} This defense requires access to the model and a small set of correctly labeled samples to reconstruct the backdoor trigger, if present. It assumes that triggers are small (to avoid detection) and finds the minimum perturbation to an input that causes misclassification to each class. Once it has candidate triggers for each class, it measures their $l_1$ norms and computes the Median Absolute Deviation (MAD) over them. MAD estimates what the standard deviation for normally distributed data would be if there were no outliers present. Neural Cleanse assumes that the $l_1$ norms of untriggered classes are normally distributed, and identifies a class as triggered if it falls outside an interval of $2\cdot$MAD around the median. The multiplier of 2 is a threshold on the anomaly index, which describes how many MADs away a sample is; 2 is chosen as it represents the 95\% confidence interval.

\noindent\textbf{TABOR~\cite{guo2020tabor}.} This defense is similar to Neural Cleanse, but replaces the $l_1$ norm with a new metric for trigger quality to reduce the number of false positive and `incorrect' triggers. TABOR assumes that a `correct' trigger is a geometric shape or symbol located in the corner of the image. 
When searching for triggers, TABOR regularizes over the largeness, scatteredness, and smoothness of triggers: in short, it penalizes reconstructed triggers that occlude important areas of the image. This enables TABOR to identify the `correct' triggers, which is followed by MAD to find the backdoored class.

\looseness=-1
\noindent\textbf{Neuron Inspect~\cite{huang2019neuroninspect}.} This defense assumes that backdoors are detectable from explanation heatmaps computed on the backdoored network. Heatmaps are created for each class by taking the gradient of the class logit with respect to the input, revealing which areas of the image were most important for predicting a given class. Next, Neuron Inspect extracts features such as sparseness, smoothness, and persistence from these maps. Assuming that the backdoored class heatmaps will be similar across all samples, the backdoored class can be identified by performing outlier detection over each class with MAD.

\subsection{Threat Model}\label{ssec:threat_model}

\looseness=-1
We now formalize our operational setting and assumptions about the training-level adversaries' capabilities. 
In the outsourced training setting, the client (\ie potential victim), submits the training set, training procedure, model architecture, and hyper-parameters to a potentially malicious ML-as-a-service cloud provider. 
The victim faces adversaries who can (a) use as much backdoored data in training as they want as they do not need to reveal that they altered the dataset, and (b) manipulate the training process at will to use stealthier backdoor attacks. For example, this adversary is capable of using the BppAttack~\cite{wangBppAttack2022} and Wasserstein~\cite{khoaWasserstein} attacks, the latter of which can achieve a near-perfect success rate without impacting performance on clean data. To the best of our knowledge, there is not yet a defense against it. Note that training-level adversaries with control over the training data and procedure are not new, but were introduced with BadNets~\cite{guBadnets}. However, these adversaries are rarely studied in works proposing backdoor defenses. 

\looseness=-1
A training-level adversary makes several previous defenses either unsuitable or fallible under our threat model. 
The victim cannot use defenses like Spectral Signature~\cite{TranSpectral} or Activation Clustering~\cite{chenActivation} due to their lack of access to backdoored data. While defenses such as Neural Cleanse, Neural Inspect, and TABOR are applicable to the threat model, they may be evaded if the adversary manipulates training algorithm properties. In \Cref{ssec:existing_defenses}, we empirically validate that 
Neural Cleanse, the most commonly studied of the three, is overcome if the adversary lowers the learning rate used when training on backdoored data. Additionally, some of these defenses are out-of-date with more recent attacks and can be thwarted by \eg WaNet, BppAttack, Adaptive Blend, and Wasserstein attack.

Realizing the difficulty of backdoor detection under this threat model, we make the following key assumption in this work: the victims, or clients, who outsource their model's training, have access to $n$ cloud providers, and not all $n$ are malicious. {We denote the lower bound of the proportion of benign servers with $r$, meaning there are at least $r\cdot n$ benign servers and at most $(1-r)\cdot n$ malicious servers.} We assume the servers do not collude, and that the malicious servers do not use identical backdoors or hyperparameters. This is a similar assumption to standard differential testing~\cite{mckeeman1998differential} which assumes that several comparable programs exist and that most are bug-free.
We also allow the client to submit parts of the training run, called \emph{sub-runs} to the other servers. 
Thus, our access to additional training servers, $rn$ of which are benign, implicitly constrains the space of backdoors an adversary can use to attack our defense. 
Unlike previous defenses which search for anomalies in the labeling behaviors over the classes, we instead identify the backdoored model behaviors over a distribution of models. This relies on the reasonable assumption that benign models differ only due to training stochasticity. While backdoor attacks each differ in their effect on the backdoored class of a model, all backdoor attacks, by definition, alter the global behaviour of the model. Our method identifies this, likely making it more capable against novel attacks. 
This is the first threat model, to the best of our knowledge, to include outsourcing training to multiple servers; a reasonable and practical model that enables a computationally limited victim to detect even the stealthiest of backdoors.

\section{\name: Our Proposed Method}
\label{sec:method}

\begin{figure}[t]
    \centering
    \includegraphics[width=\linewidth]{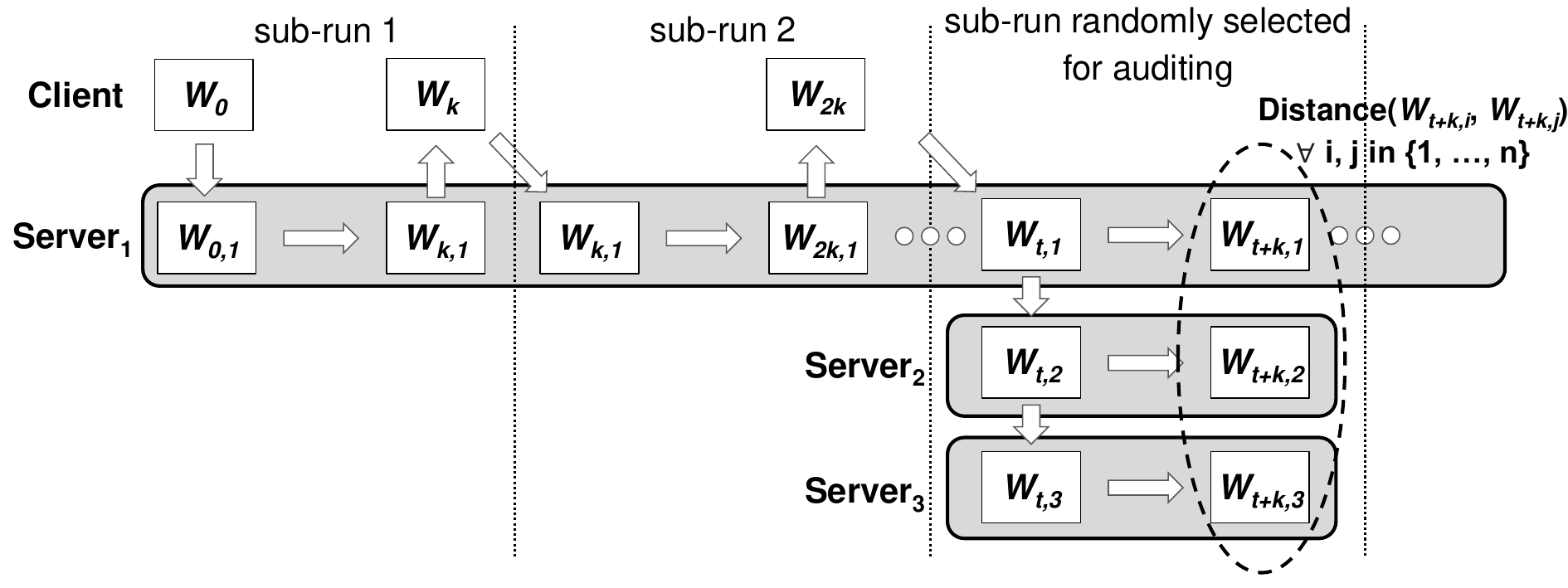}
    \caption{\looseness=-1 An illustration of \name, our proposed approach for detecting backdoor insertion. The client submits a training task to the primary server, denoted by Server$_1$, and downloads the intermediate model checkpoint after every sub-run, each of which consists of $k$ training updates. At the beginning of an arbitrary sub-run at training step $t$, the training task and model parameters $W_t$ are probabilistically submitted to the $n$ servers the client has access to ($n=3$ in this diagram). Afterward, the client collects $W_{t+k,1} \dots W_{t+k,n}$ returned by the servers, and computes the pairwise distances among them based on a chosen distance metric, or equivalently, $distance(W_{t+k,i}, W_{t+k,j})$ $\forall i,j \in [1 \dots n]$ and $i < j$. It is expected that the distances among models returned by benign servers form a cluster since they are obtained by running the same training process. We thus verify if pairwise distances from the primary server to other clean models fall in this cluster to detect  primary servers that are acting maliciously.
    \vspace{-5mm}
}
    \label{fig:method}
\end{figure}

This section introduces our backdoor detection method, \name (Replicate Training To Detect), which is capable of detecting backdoors inserted by training-level adversaries in the outsourced training setting. Before introducing our entire method, we first motivate our key design choice, \textit{replicated training of sub-runs}, in Sections~\ref{ssec:intuition}, and demonstrate how this provides a distribution over model updates. In~\Cref{ssec:distribution}, we show how we can identify malicious servers by comparing the updates in this distribution using metrics that measure the distance between models. We use statistical tests to complete our backdoor detection process. In~\Cref{ssec:prob_cost} we overview the probabilistic guarantee given by \name and analyze the associated costs.

\subsection{Outsourced Replicated Training}
\label{ssec:intuition}
\looseness=-1
To begin outsourced training, the client chooses a model architecture, a training algorithm (\eg a variant of stochastic gradient descent like vanilla SGD or Adam), and a dataset for the server to run. As part of the training algorithm, the client also specifies hyperparameters such as the number of training steps $k$
, the learning rate, and the mini-batch size. We also allow the client to choose the weight initialization for the model parameters that the servers begin training from. We believe this is a realistic assumption since it is common practice to fine-tune a pre-trained foundation model. Then the client submits a training job with these specifications to the servers, and we call this $k$-step training process a \textit{sub-run}. After the sub-run is completed, the client downloads the model weights and the next sub-run begins. If a model needs $T$ training steps to converge, then there will be $T/k$ sub-runs in total.%

With this procedural understanding of outsourced training, we provide some intuition on the plausibility of detecting backdoors in this setting. First, we want to emphasize the following fact: simultaneously learning a backdoor and a primary task (\eg classification) differs from, and is harder than, only learning the primary task. Thus, model updates for the backdoored task will differ from and be drawn from a different distribution compared to updates computed for the primary learning task alone.\footnote{While it is possible to backdoor a model by finding a particular re-ordering of clean data points in each minibatch~\cite{shumailov2021manipulating}, %
the resulting update will remain out-of-distribution compared to the updates corresponding to a randomly-sampled minibatch.} %
Hence, the key challenge when detecting backdoors in our setting is modelling the distribution of benign updates. If this were solved, backdoor detection would be nothing more than a statistical test or outlier detection over this distribution. This is where replicated training proves useful: it enables us to obtain a distribution of benign model updates, provided a multiplicity of benign servers.

\looseness=-1
As mentioned in Section~\ref{ssec:threat_model}, we rely on the assumption of having access to multiple servers available for outsourced training. In a similar manner to differential testing~\cite{mckeeman1998differential}, our approach, \name, obtains replicated model updates for some of the sub-runs from the pool of the available $n$ servers. First, the client chooses one server as the \textit{primary server}, indexed as server 1, which completes the entire training run. The client would simply upload their training specifications to it, and download model weights after each sub-run, \ie every $k$ steps
. The weights at the start and end of the sub-run are denoted by $W_t$ and $W_{t+k}$, as the sub-run occurs from step $t$ to $t+k$. 

\looseness=-1
The client also chooses $m$, the number of sub-runs that undergo replicated training. Based on $m$, the client may decide to perform replicated training for a given sub-run. If so, the client submits their training specifications to the other $n-1$ servers, indexed by $i\in\{2..n\}$. After training for $k$ steps, these $n$ servers (including the primary server) return models with weights denoted as $W_{t+k,1}, W_{t+k,2} \dots W_{t+k,n}$ respectively.  Note that here the subscripts correspond to the \textit{expected} number of training steps performed by the server to complete the assigned sub-run. If a server $i$ acts maliciously, perhaps by training for a longer or shorter time, the model parameters $W_{t+k,i}$ returned by the server may have resulted from training with more or fewer than $k$ steps. We show an example of this process in~\Cref{fig:method}, where the client replicates training to three servers (including the primary server) at the third sub-run. 

One of the most important parameters for the client to set is $k$, the number of training steps in the sub-run. This parameter impacts the variance of the distribution of model updates given by $W_{t+k,1}, W_{t+k,2} \dots W_{t+k,n}$, as training stochasticity scales with the number of steps taken. The client can decrease the variance of the benign server distribution by lowering $k$; in the extreme case where $k=1$, this should completely eliminate the variance outside of small differences caused by the software and hardware stacks. Consider a malicious server with access to the distribution of model weights returned by benign servers. In order to evade detection, the adversary would want to return a model update that falls within the variance of the benign distribution. If this variance is very small, the adversary faces a difficult challenge as their backdoor update is out-of-distribution compared to the primary training task (recall~\Cref{ssec:intuition}). In addition, $k$ negotiates a trade-off between the benign update distribution variance (and therefore the ease of detection) and the communication cost incurred from uploading and downloading model weights. 

The parameter $m$, the number of replicated sub-runs, also manages the cost of replicated training, but should be set considering the number of sub-runs that a malicious server chooses to backdoor. In extreme cases, a malicious server might only backdoor during one sub-run to evade detection. This is why our method focuses on whether a given sub-run underwent a backdoor attack instead of attempting to identify a malicious server based on a history of malicious updates.

While replicating training can obtain a distribution for model updates, characterizing these distributions is nontrivial due to the high-dimensionality of model updates. If $k$ is relatively large, it may also be challenging to distinguish between updates arising from adversarial behavior and benign training stochasticity. Next, we overview a method to manage these challenges, and describe the core components of \name.

\subsection{Model Distances}
\label{ssec:distribution}

\looseness=-1
From the distribution of model updates, \name uses model distance metrics to reduce dimensionality and identify any malicious model updates. 
To reduce this dimensionality and identify backdoored model updates, \name uses model distance metrics to quantify the distances between models.  
Following \Cref{alg:main}, the client begins by computing the distance between every pair of returned models:  $distance(W_{t+k,i}, W_{t+k,j})$ $\forall i,j \in [1 \dots n]$ and $i < j$. These $\binom{n}{2}$ pairwise distances over the $n$ models result in three groupings of servers for comparison: benign v.s. benign, benign v.s. malicious, and malicious v.s. malicious. 
Intuitively, any two benignly trained models with similar training configurations starting from the same initialization should not significantly differ from each other. In contrast, since we assume non-colluding malicious servers that each use a unique backdoor, we expect the backdoored model updates to differ greatly from benign model updates and from the other backdoored model updates of different adversaries.
Therefore, of the three groups of pairwise distances, we hypothesize (and later validate in~\Cref{ssec:detection_eval}) that the benign v.s. benign group has the least variance. If we consider each group as a cluster, the benign v.s. benign cluster should contain roughly ${ r n \choose 2}$ of the pairwise distances, where r is the proportion of benign servers (recall \Cref{ssec:threat_model}). 

\looseness=-1
To find this cluster, we sort the pairwise distances by their magnitude and begin an iterative procedure to identify the cluster with the smallest variance. We iterate over integer values of $l$: at each iteration identifying the cluster $c_l$ containing the $l^{th}$ to $(l+ {rn \choose 2})^{th}$ sorted distances and finding its variance. Eventually, this returns the cluster with the smallest variance, \ie $c = argmin_{c_l} \sigma^2(c_l)$. To determine whether the primary server is benign, we consider the set of pairwise distances between it and the other servers, $\{distance(W_{t+k,1}, W_{t+k,j}) |  j \in [2 \dots n] \}$, which we refer to as the ``primary-to-other distribution.'' Given that there are at least $r n$ benign servers, if the primary server is benign, then $r n - 1$ pairwise distances in this set should be between two clean models. In other words, for the primary server to be benign, there must be $rn-1$ primary-to-other distances from the same distribution as the minimum-variance cluster distances $c$. Since the distances are scalars, we use a Kolmogorov-Smirnov (KS) test with a significance level of 0.01 to decide if this null hypothesis (that the primary server is benign) should be rejected.
If the hypothesis passes, we claim the primary server is benign; the client can then choose to run the next sub-run on the primary server or subject the next sub-run to \name --- thus repeating our procedure. It is noteworthy to mention that while we focus on verifying whether the primary server backdoored the sub-run, our method can also detect whether any of the servers performed a backdoor attack. To do so, the client need only perform the KS test using the same cluster $c$ and the set of distances between the model returned by the server of interest and the other models.

\noindent\textbf{Distance metric} The final consideration when using model distance metrics in \name is choosing the appropriate metric. A typical first choice may be flattening the model weights and taking the cosine distance between them to get the \textit{parameter-space distance}~\cite{jia2021proof} (potentially with parameters aligned first since some model layers are invariant to transformation). While being computationally efficient, we find in~\Cref{ssec:adaptive_adv} that \name using the parameter-space distance can be fooled by an adaptive adversary optimizing such that the backdoored model parameters are close to those of a surrogate clean model. 
Therefore, the client should choose a distance metric that captures the behavior of models, especially because the backdoor task induces a new behavior that is distinct from the primary training task. Adequate metrics include the cosine distance between the model's outputs on a set of clean testing data, a.k.a., \textit{output-space distance}, the Zest distance, and the Centered Kernel Alignment (CKA) similarity~\cite{kornblith2019cka}. 
Zest measures similarities/differences between models by leveraging a set of reference data points. 
For each point, a small dataset is generated by randomly masking out segments of the points and having the model of interest label this masked sample. Next, a linear model is trained on the masked samples to capture the model's behavior locally around each reference data point (\ie the model learns what class label each segment corresponds to). Concatenating the weight vectors of these linear models gives a signature for the model, where the distance between these signatures represents the difference between the global behaviors of the models. Refer to~\Cref{app:model_distances} for additional information on Zest.
CKA similarity compares the activations that two models output at one of their hidden layers; we use the linear variant of CKA and compare the activations resulting from the final layer. 
While we will evaluate all these metrics discussed so far in \Cref{ssec:eval-ablation}, we use Zest for \name in most of experiments unless otherwise specified. Zest requires less data to function than output-space distance, is more computationally efficient than CKA similarity, and, as we empirically demonstrate in~\Cref{ssec:adaptive_adv}, is nontrivial to fool by the adaptive adversaries we consider.

\begin{algorithm}[t]
    \begin{algorithmic}
    \State \textit{AllDist, PrimaryDist, Cluster} $= [], [], []$
    \For {$i=1,2,\ldots,n$}
        \State $W_{t+k,i}$ = Outsourced training$(W_t, $\textit{server}$_{i}, $\textit{steps}$=k)$
    \EndFor

    \For {$i=1,2,\ldots,n$}
        \If {$i>1$}
            \State \textit{PrimaryDist}.append(\textit{distance}$(W_{t+k,1}, W_{t+k,i})$)
        \EndIf
        \For {$j=i+1,i+2,\ldots,n$}
            \State \textit{AllDist}.append(\textit{distance}$(W_{t+k,i}, W_{t+k,j})$)
        \EndFor
    \EndFor

    \State Sort \textit{AllDist, PrimaryDist}
    \For {$l=1,2,\ldots$}
        \State \textit{Cluster}.append(\textit{AllDist}$[l:l+{ rn \choose 2}]$) 
    \EndFor
    \State $c = argmin \ \sigma^2$(\textit{Cluster}$[l])$ \Comment{Cluster with the smallest $\sigma^2$}

    \For {$o=1,2,\ldots$}
        \If {KS test($c$, \textit{PrimaryDist}$[o:o+r n - 1]$) $< 0.01$}
            \State The Primary server is benign.
        \EndIf
    \EndFor
    \end{algorithmic}
    \caption{\name (for Sub-run at step $t$) }
    \label{alg:main}
\end{algorithm}

\subsection{Probabilistic Guarantee and Cost Analysis} \label{ssec:prob_cost}
\looseness=-1
Lastly, it is worth noting that our proposed backdoor detection approach provides a probabilistic guarantee. If we assume a powerful adversary is able to inject a successful backdoor by manipulating just a single step of training, the chance of that backdoor insertion being detected is $m \cdot k / T$, where $m$ is the number of sub-runs that the client verifies, $k$ is the length of the sub-runs, and $T$ is the number of total training steps. Note that this is the worst-case analysis. In practice, backdoors are subject to catastrophic forgetting~\cite{liu2017neuraltrojan} and could be forgotten if the adversary does not reinforce them during training.
Furthermore, the primary server is unaware of the total number of sub-runs, so it is challenging for an adversarial primary server to insert a backdoor during the final sub-runs (where it is less likely to be forgotten). Put together, this makes it harder for an adversary to evade detection by \name, even though the guarantee is probabilistic. 
Moreover, if the client is willing to increase them, \ie $m \cdot k \to T$, the probability of a malicious server evading detection decreases to approach $0$. The expected additional cost incurred by outsourcing training to multiple servers is equivalent to conducting $m \cdot k \cdot (n-1)$ additional training steps.

\begin{figure}
    \centering
    \includegraphics[width=0.95\linewidth]{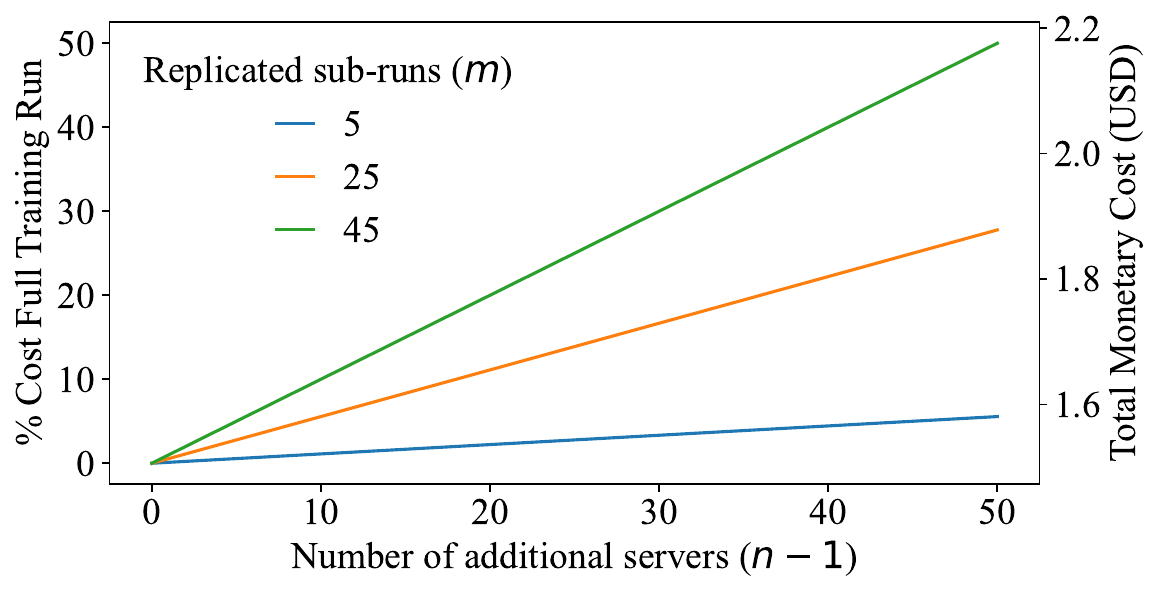}
    \caption{
    \looseness=-1
    Cost overhead incurred by replicated training in \name as a function of the number of additional servers across different numbers of replicated sub-runs, $m$. We consider a training job on CIFAR10 containing 200 epochs and the sub-run length $k$ is set to 2000 steps such that there are 45 subruns in total. The cost is presented as a percentage of the number of training steps in a full training run on the left y-axis, and as monetary cost on the right y-axis according to AWS SageMaker using the same compute as described in \Cref{ssec:exp_platform}. Note that the right y-axis is simply a rescaled version of the left axis.
    \vspace{-3mm}
    }
    \label{fig:cost_analysis}
\end{figure}

\looseness=-1
The parameters $m$ and $k$ impact the probabilistic guarantee and the expected additional cost incurred by outsourcing training to multiple servers, which is equivalent to conducting $m \cdot k \cdot (n-1)$ additional training steps.
The other most significant computational cost of \name is for computing the model distance metric for the $\binom{n}{2}$ pairs of models, which we quantify in terms of the equivalent number of training steps. For simplicity of comparison, we assume that the output-space distance, Zest, and CKA similarity use $D$ mini-batches of data. The main expense of these metrics is their forward propagation, which would cost $\frac{D}{3} \cdot \binom{n}{2} \cdot m$ training steps\footnote{In algorithms using gradient descent, backward propagation is approximately twice as expensive as forward propagation}. The parameter-space distance only needs to compute a total of $\binom{n}{2} \cdot m$ cosine distance computations, each of which has a similar number of floating point operations to a forward pass. Thus, the client-side cost is similar to conducting $\frac{1}{3} \cdot \binom{n}{2} \cdot m$ training steps. \Cref{fig:cost_analysis} shows the total additional cost of \name as a percentage of the number of training steps of training the model normally. It also shows the monetary cost of outsourcing training according to the cost of using AWS SageMaker.\footnote{See AWS SageMaker pricing here: \url{https://aws.amazon.com/sagemaker/pricing/?did=ap_card&trk=ap_card}.}

\section{Evaluation}\label{sec:evaluation}
\looseness=-1
Now that we have described our method, \name, we empirically validate its effectiveness against existing and adaptive backdoor attacks that are representative in the outsourced training setting. We first describe our experimental setup, then present a series of experiments designed to answer the following questions: (1) How do existing defenses fare against attacks from the training-level adversaries present in our outsourced training threat model? (2) Is \name able to effectively detect backdoor attacks? (3) Can \name be bypassed by adaptive adversaries utilizing knowledge of our method to stealthily insert a backdoor? and (4) Are there training settings that \name does not apply to? Based on these experiments, we find that:

\begin{enumerate}[leftmargin=*,noitemsep,topsep=0pt]
    \item Existing defenses are vulnerable in our threat model. A training-level adversary can use a simple modification to the BadNets attack to fool existing detectors by decreasing the backdoor signal. Details are presented in~\Cref{ssec:existing_defenses}.
    \item \name is able to detect backdoor attacks with 99.6\% accuracy. We find that we can effectively control the amount of training stochasticity using $k$, allowing Zest to compare the global behavior of models and identify a concentrated cluster of benign model updates. Additionally, \name is capable of detecting every backdoor we evaluate, including the Wasserstein attack, which to the best of our knowledge has never been defended against so far. Details are presented in~\Cref{ssec:detection_eval}.
    \item We evaluate \name against adaptive attacks that attempt to optimize such that the distance between their backdoored models and clean surrogate models falls within the distribution of pairwise distances between clean models. We find that the adaptive adversary is unable to achieve this even with full knowledge of the reference data  used to evaluate the Zest distance. Details are presented in~\Cref{ssec:adaptive_adv}.
    \item We conduct an ablation study to show that \name consistently performs well with a range of training settings. 
\end{enumerate}

\subsection{Implementation Details} 
We primarily validate \name in supervised learning for image classification, which is the focus of most existing backdoor attacks.
Our experimental results were obtained by repeating the experiments at least five times, where possible. 
We use a ResNet-20~\cite{HeResnet} architecture trained on the CIFAR-10 dataset~\cite{cifar10} for 200 epochs and a VGG-11~\cite{vgg} architecture trained on the GTSRB dataset~\cite{gtsrb} for 60 epochs, which achieve average test accuracies of $91.77 (\pm 0.17) \%$ and $92.52 (\pm 0.59) \%$ respectively across five runs of benign training (\ie no backdoor inserted). We also test \name on the ImageNet dataset~\cite{imagenet} by continuing training on a pre-trained VGG-19~\cite{vgg} model, with test accuracy of $72.39\%$.
Following the threat model described in \Cref{ssec:threat_model}, a primary training server, which can be either benign or malicious, performs a complete training process consisting of $T$ training steps, each corresponding to a gradient update computed on a mini-batch of data points. 
We implement this training run in multiple sub-runs, each of $k<T$ steps. Refer to \Cref{ssec:exp_platform} for details on our compute and experimental platform.

During training, the client selects sub-runs of this training process and asks the other $n-1$ training servers to recompute these sub-runs. 
To simulate different cloud providers operating each of these $n-1$ servers, we implement each server on a different physical machine and vary the random seed used by each.
Recall that our threat model includes training-level adversaries that may arbitrarily manipulate the training data and training process, as they need only provide the model weights at the end of the sub-run. For example, a malicious server might manipulate the backdoor learning rate or might use extra training steps to insert a backdoor.  From the client's perspective, outsourced training is a black box, the client has no reliable insight into the training process other than the weights of the model trained by it. 

\looseness=-1
To experiment with malicious servers, we implement eight backdoor attacks based on variants of the five strategies described in \Cref{ssec:background_backdoor}: BadNets~\cite{guBadnets}, WaNet~\cite{nguyen2021wanet}, BppAttack~\cite{wangBppAttack2022}, Adaptive Blend~\cite{qi2023revisiting}, and Wasserstein Backdoor~\cite{khoaWasserstein}. We use three differently-sized triggers for BadNets, from smallest to largest they are a white square, white stripe, and RGB flag, as shown in~\Cref{fig:trigger} in Appendix~\ref{app:additional_figures}. We include both variants of the Wasserstein backdoor which differ in the architecture (U-Net or autoencoder) used to generate the triggers. In addition to the eight malicious servers corresponding to these attacks, we also include eight benign servers who perform the training faithfully. 
{Note that by considering $50\%$ of servers being malicious, we are considering a much harder scenario than when only one or two servers are malicious, as the difficulty of modeling distributions of benign updates is easier when there are more benign servers.}
Aside from the ablation studies, the sub-run length $k$ is set to the number of steps in five training epochs (we round to 2000 for CIFAR-10 and 1500 for GTSRB), and we consider a training sub-run occurring halfway through the total training run (\ie after 100 epochs for CIFAR-10 and after 30 epochs for GTSRB).

\looseness=-1
Note that the goal of this work is to detect backdoor attacks deployed by malicious training servers. In particular, we strive to detect any non-trivial backdoor success, \ie data points that contain a trigger are predicted as the class chosen by the adversary with a probability higher than random guessing (\eg $10\%$ for CIFAR-10).
For example, the attack success rate of some backdoors comes at the cost of reduced clean data accuracy. We configure such attacks so they have similar accuracy to clean models but attack success that is only slightly better than random guessing. Otherwise, it would be trivial for the client to detect accuracy-lowering malicious activity as most of the replicated model updates would improve the accuracy.
This choice makes the backdoors harder to detect, which is a strength of our evaluation from a defense perspective. This is further confirmed in~\Cref{ssec:eval-ablation} where we ablate over the attack success rate and show that \name is not sensitive to it.

\subsection{Evaluating Prior Defenses}\label{ssec:existing_defenses}
\begin{figure}
    \centering
    \includegraphics[width=0.95\linewidth]{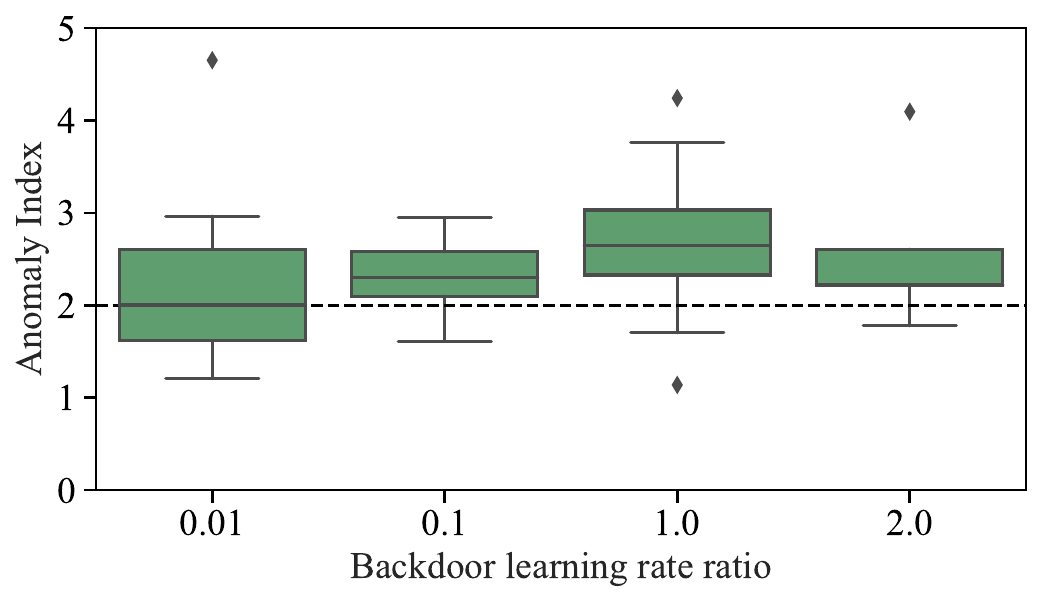}
    \caption{The anomaly index of Neural Cleanse~\cite{Wang2019Neural} falls below 2, the backdoor detection threshold, when a malicious server lowers the learning rate for backdoored data. We represent this as the ratio between the clean and backdoor learning rates. We report the 95\% confidence interval taken over ten random seeds corresponding to ten backdoored models per learning rate ratio, all on CIFAR10. \vspace{-3mm}}
    \label{fig:lr_ratio_existing_defenses}
\end{figure}

\looseness=-1
Here we show how existing backdoor attacks can be thwarted, even on the attacks they were designed to predict, due to our threat model. For example, Neural Cleanse~\cite{Wang2019Neural} was originally designed to detect BadNets triggers. Recall that Neural Cleanse uses Median Absolute Deviation (MAD), which gives an anomaly index larger than a threshold set to 2 for classes where backdoors are suspected. In ~\Cref{fig:lr_ratio_existing_defenses} we show that the anomaly index decreases and the defense fails if a training-level adversary simply lowers the learning rate used on the backdoored data. Lowering the learning rate lowers the strength of the backdoor attack for the target class, decreasing how anomalous that class is compared to the others. We show that half of the backdoored models would not be detected by Neural Cleanse when the backdoor rate is 0.01 times the clean learning rate. Hence, we conclude that existing defenses are insufficient in the outsourced training paradigm. For the sake of completeness, in~\Cref{ssec:ed_outsourced} we present an alternative version of \name that adapts existing defenses to better fit our threat model, though it has the same limitations as already described here.

\subsection{Backdoor Detection Evaluation}
\label{ssec:detection_eval}

\begin{figure}[t!]
    \centering
    \begin{subfigure}[b]{\linewidth}
        \includegraphics[width=0.95\linewidth]{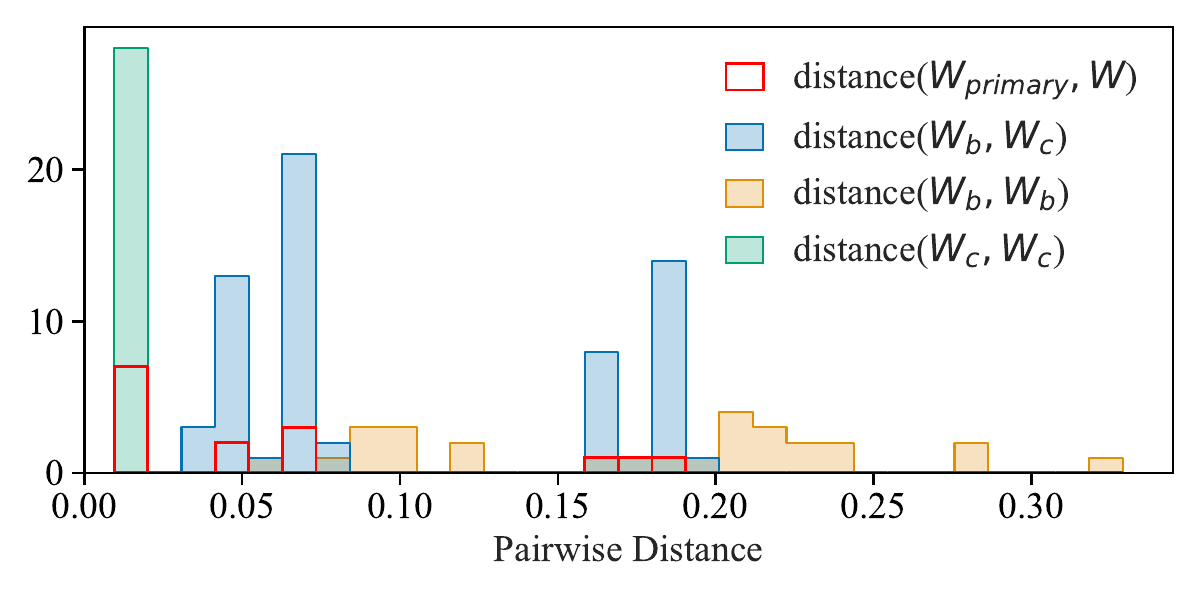}
        \caption{Pairwise Distance}\label{subfig:main_benign}
    \end{subfigure}
    \begin{subfigure}[b]{\linewidth}
        \includegraphics[width=0.95\linewidth]{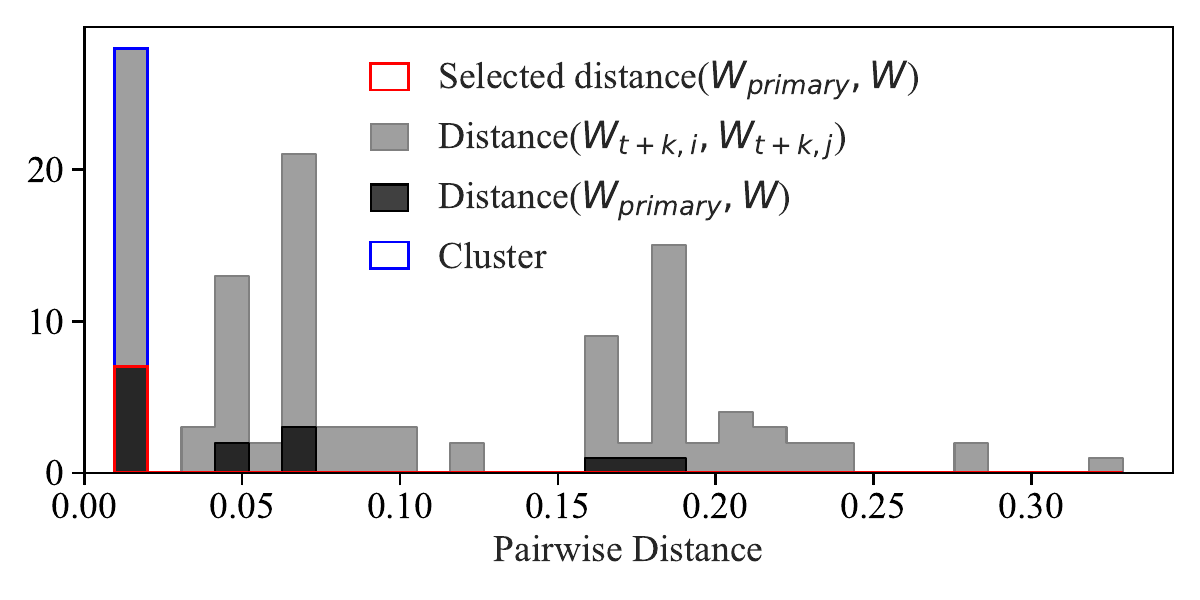}
        \caption{Client's View}\label{subfig1:cluster_benign}
    \end{subfigure}
    \caption{
        Histogram of pairwise distances among 16 ResNet models on CIFAR10 from a sub-run with the length of five epochs, {where the x-axis represents the model distance, and the y-axis is for the count of pairwise distances within each distance range.}
        Eight models are trained by benign servers (denoted by $W_c$) and the other eight are trained by malicious servers using different backdoor strategies or triggers (denoted by $W_b$). 
        In \Cref{subfig:main_benign}, different colors are used to represent the categories that each of the model pairs belongs to when computing pairwise distances. Note that distance$(W_c,W_c)$ all fall into a single bin whereas distance$(W_b,W_c)$ and distance$(W_b,W_b)$ have larger variances.
        \Cref{subfig1:cluster_benign} is what the clients observe. The distances from the model computed by the primary server to all other models are represented by the bins distance$(W_{primary},W)$, in which $r \cdot n - 1$ instances are selected to approximate distance$(W_{primary},W_c)$ as circled in red. They overlap with the cluster circled in blue approximating distance$(W_c,W_c)$, meaning the primary server is benign.
    \vspace{-3mm}
    }
    \label{fig:main}
\end{figure}

\looseness=-1
We now evaluate \name. Using our replicated training setting, the client may outsource sub-runs multiple times and obtain a distribution of model updates computed by different servers. 
The client then computes pairwise model distances between every pair of models returned by the servers.
In these results, we use \name with the Zest distance, though an ablation study on different distance metrics in \Cref{ssec:eval-ablation} shows that all distance metrics we consider allow \name to identify the benign cluster. These pairwise distances for ResNet-20 trained on CIFAR-10 are plotted in \Cref{fig:main} (see \Cref{fig:main_gtsrb} and \Cref{fig:imagenet} in Appendix~\ref{app:additional_figures} for VGG-11 trained on GTSRB and VGG-19 trained on imagenet~\cite{imagenet} respectively). We use blue to show benign pairs (distance$(W_c,W_c)$), green for a benign vs. backdoored (distance$(W_b,W_c)$), and orange for backdoored pairs (distance$(W_b,W_b)$). We also highlight (in red) the distances from the model trained by the primary server to models trained by other servers. 
It can be seen that distance$(W_b,W_b)$ and distance$(W_b,W_c)$ vary over a large range of distance values, likely due to different malicious servers using distinct backdoor strategies and triggers which correspond to diverse local optima noted by Zest. In fact, this variance is likely an underestimation since (a) we implicitly assume some shared strategies across the adversaries, such as that they will try to be stealthy and not harm the clean data accuracy, and (b) there currently exists a multitude of other backdoor strategies.

\looseness=-1
While the Zest distance between pairs of models that include a backdoored model have large variances and ranges, the distances between two clean models (distance$(W_c,W_c)$) cluster to the same histogram bin. This is because the only differences between benign model updates arises due to randomness incurred in training (\eg random data augmentation, random data ordering, or hardware-level noise). While there potentially may be large amounts of stochasticity accumulating over five epochs ($k$), the training data is unaltered, and the behavior of benign models remains similar. Because the models share their global behavior, the Zest differences between them are small.

Note that \Cref{subfig:main_benign} is only meant to help with the reader's understanding. In reality, the client only observes a histogram similar to \Cref{subfig1:cluster_benign}. \name first approximates the distribution of distance$(W_c,W_c)$ by finding the cluster of $\binom{r \cdot n}{2}$ pairwise distances with the smallest variance, as shown by the histogram bins circled in blue. It can be seen this approximation is accurate since the cluster is identical to the bin of distance$(W_c,W_c)$ in \Cref{subfig:main_benign}. We then form a null hypothesis assuming the primary server is benign, or equivalently, at least $(r\cdot n) - 1$ of distance$(W_{primary},W)$ (circled in red) are from the same distribution as the approximated benign cluster. A Kolmogorov–Smirnov (KS) test with a significance level of 0.01 is then used to decide whether this hypothesis should be rejected. In this case, the null hypothesis is not rejected since the primary server is indeed benign. The scenario of a malicious primary server is shown in \Cref{fig:main_malicious} in Appendix~\ref{app:additional_figures}.

\looseness=-1
As mentioned in \Cref{ssec:distribution}, \name can also be applied to detect if any other servers involved in the replicated training are malicious. We show the results of doing so to all the 16 servers in \Cref{table:ks_test}, alongside 14 other runs with different settings such as datasets and training hyperparameters. Only 1 of the 240 servers is misclassified by \name, which happens in an earlier stage of training. Thus we do not consider it worrisome as explained in \Cref{ssec:eval-ablation}. Another important observation is that if the model is backdoored before the sub-run, \name does not falsely blame servers acting benignly in the sub-run where \name is applied, as shown in the third row of the table.

\begin{table}[t]
\vspace{2mm}
\centering
\normalsize
\scalebox{0.9}{
\begin{tabular}{l | c | c c c}
\toprule
Model & Server & p-value & max & min \\ 
\midrule
\midrule

\multirow{2}{*}{Clean} & Benign & $0.689 (\pm 0.303)$ & 1.000 & 0.002 \\
 & Malicious & $< 10^{-3}$ & $< 10^{-3}$ & $< 10^{-3}$\\

\hline

\multirow{2}{*}{Backdoored} & Benign & $0.653 (\pm 0.308)$ & 0.988 & 0.148\\
 & Malicious & $< 10^{-3}$ & $< 10^{-3}$ & $< 10^{-3}$ \\

\bottomrule
\end{tabular}
}
\caption{P-values of Kolmogorov–Smirnov (KS) tests. The initial model checkpoint (up to the epoch indicated by step $t$) is trained by either a benign or malicious primary server, thus the model could be clean or backdoored (first column). Then each model checkpoint is submitted to eight benign servers and eight malicious servers for sub-run training (second column). We repeat the experiments 15 times with varying client or server-side settings (\eg datasets, training hyperparameters, etc.). Following \name, we apply a KS test to each server with the null hypothesis that the server is benign and report the p-values in the last three columns. 
Note that the minimal p-value in the third row is below the significance level of 0.01. In fact, it is the only incorrect detection among the 240 servers in the 15 runs we tested, leading to an overall detection accuracy of $99.6\%$. 
This single error can be corrected at no cost by lowering the significance level to 0.001.
\vspace{-3mm}
}
\label{table:ks_test}
\end{table} 

\looseness=-1
\para{\name against Language Backdoors}
We also test \name in the setting of language classification tasks. Specifically, we consider pretrained BERT models~\cite{bert} and finetune them on the AG News dataset~\cite{zhang2015character}. During finetuning, there are three benign servers and two malicious servers using BadNets (\ie attaching a few special words to the input) and Hidden Killer~\cite{qi2021hidden}, respectively. We found that \name was able to correctly classify all of the 5 servers  (see \Cref{fig:bert} in Appendix~\ref{app:additional_figures}), demonstrating \name's applicability outside the image domain.
However, because Zest is only applicable to classification tasks, extending \name to, for example, generative models, will require a distance metric that captures behaviors of such models. To the best of our knowledge, such a distance metric does not yet exist, and should be left to future work.

\subsection{Adaptive Adversaries}
\label{ssec:adaptive_adv}
\looseness=-1
Previously, we presented empirical results with the assumption that the adversary is unaware that \name is being used. We now relax this assumption and consider two adaptive adversaries that know \name is being used but have differing levels of knowledge: (1) adversaries that know the distance metric being implemented, and (2) adversaries with further knowledge about the reference data points that the distance metric is computed with respect to (this only applies to output-space, CKA, and Zest distance). 
We also assume the adaptive adversaries have rich computational resources such that their attacks are not restricted by cost. 
Both types of adversaries attempt to optimize the distance between their backdoored model and clean models, distance$(W_b,W_c)$, so that it falls within the distribution of distance$(W_c,W_c)$. We identify several methods the adversaries might use, and evaluate \name against these attacks. 

\para{Parameter-space Distance} 
The parameter-space distance is a Euclidean or cosine distance directly computed on model parameters, therefore an adaptive adversary may attempt to limit the deviation in parameters caused by their attack. We consider an attack where the malicious server lowers their learning rate such that the parameters of their backdoored model differ little from the parameters provided by the client at the start of the sub-run. The attack is enabled by the overparameterization of ML models, where two models with similar parameters may have distinct behaviors. We show in~\Cref{fig:distrib_parameter} that the pairwise parameter-space distances between two backdoored models or one backdoored and one clean model can form clusters with smaller variance as the learning rate decreases, and these clusters may overlap with the distances between clean models that used the client-specified training rate. This means that \name might not be able to correctly identify the benign vs. benign group, leading to inaccurate detections. Fortunately, this attack does not fool Zest or the CKA distance as they are not directly dependent on the model parameters, which is a reason why we propose to use Zest by default in \name.

\para{Zest Distance}
Recall from Section~\ref{ssec:distribution} that Zest captures differences in the model's behavior. It relies on training linear models using predictions of the target models on randomly masked samples of some reference data points. In the extreme, if a model can have identical outputs to another model with respect to every possible masked sample of the exact reference data points, then the two models will have a $0$ zest distance.
Thus our adaptive adversaries aim to encourage the backdoored model to perform as similarly as possible to a clean model in such masked samples. Note that depending on their knowledge of Zest's reference data, they either use the exact reference data or create a similar set (\eg by randomly sampling from the training dataset). The adversary then trains a clean surrogate model according to the client's instructions, and perturbs their backdoored model such that it behaves similarly to the surrogate model with respect to some randomly masked samples of the reference data points, by \eg minimizing a mean squared error (MSE) between the outputs of the two models after every time training on backdoored data.

\begin{figure}[t!]
\centering

\includegraphics[width=0.95\linewidth]{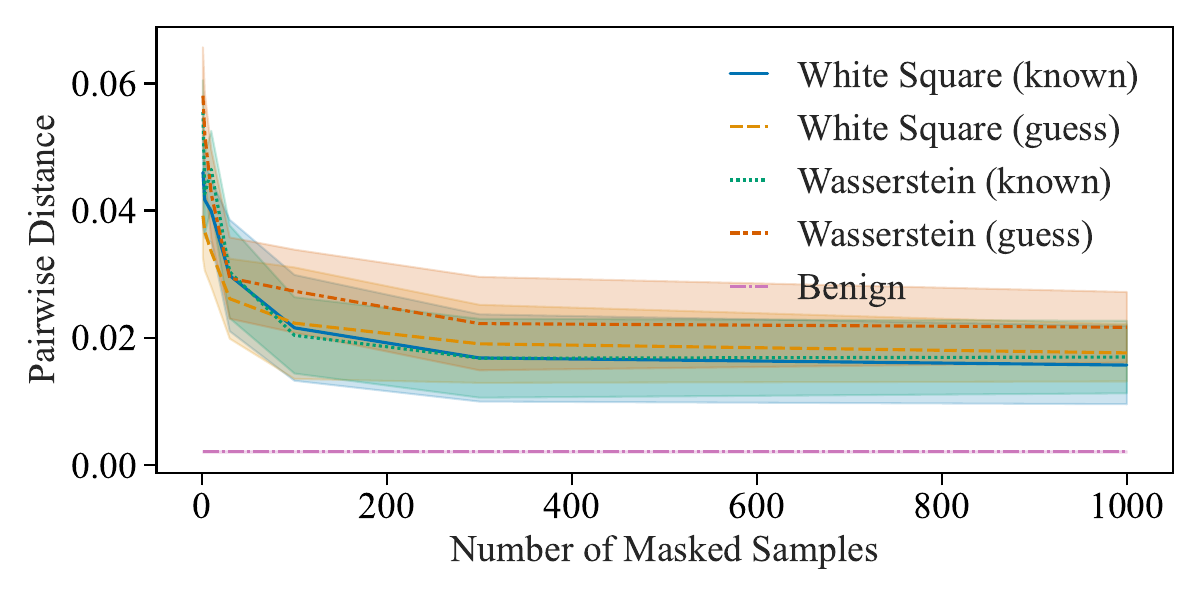}

\caption{
Pairwise Zest distances between clean and backdoored models under an adaptive attack. For reference, see the distances between two clean models (the curve labeled ``benign''). We consider four scenarios including two types of backdoor attacks and two degrees of adversarial knowledge about the Zest distance computation in \name. For each scenario, 
the plotted curve represents the best pairwise distances obtained across five runs of adaptive attack (\ie the trained $W_b$ whose distance$(W_b,W_c)$ is closest to the distribution of distance$(W_c,W_c)$.). Despite the adversary being able to decrease distance$(W_b,W_c)$ by increasing the number of masked samples, they still do not overlap with distance$(W_c,W_c)$.
\vspace{-3mm}
}
\label{fig:adaptive_samples}
\end{figure}

\begin{figure*}[t]
\centering

\subfloat[Parameter-space Distance
]
{
\includegraphics[width=0.32\linewidth]{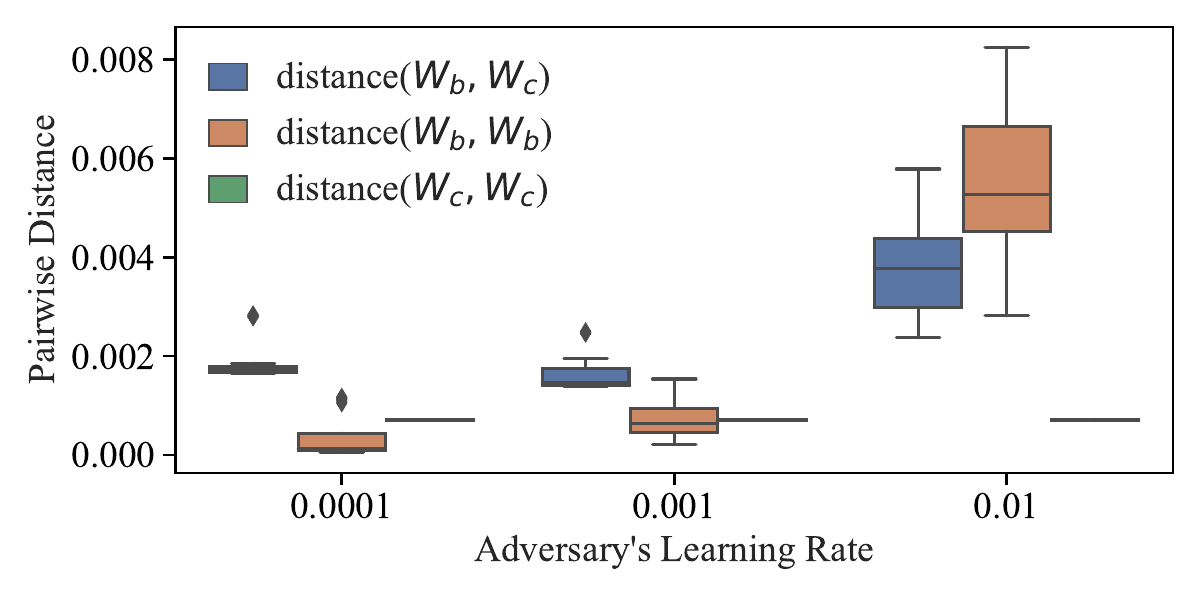}
}
\subfloat[Zest Distance
]
{
\includegraphics[width=0.32\linewidth]{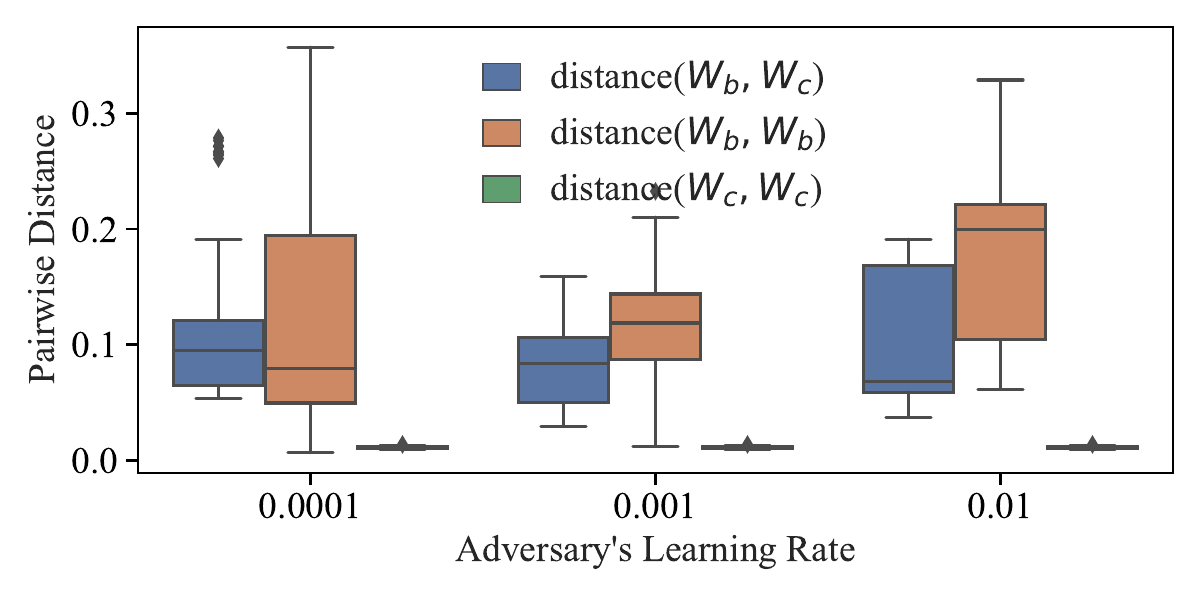}
}
\subfloat[CKA Distance
]
{
\includegraphics[width=0.32\linewidth]{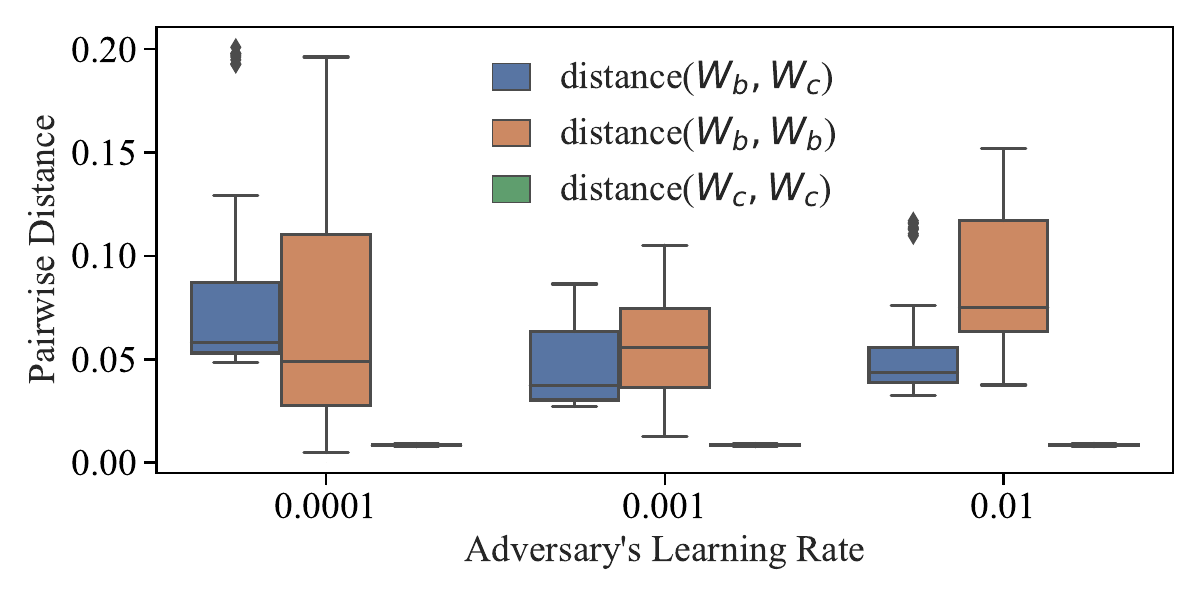}
}

\caption{
This figure shows three pairwise distances: Zest, CKA, and the parameter-space distance undergoing an attack. In this attack, the malicious server decreases the learning rate to reduce the variance of parameter-space distances between two backdoored models, or one backdoored model and one clean model. This follows from the triangle inequality: a smaller learning rate means the models are closer to the model weights the sub-run starts from, and thus the parameters are close to each other. When the learning rate is small enough, this results in distance$(W_c,W_c)$ that are indistinguishable from distance$(W_b,W_c)$ or distance$(W_b,W_b)$ when considering their respective variance. In contrast, we see that Zest and CKA distance makes \name robust against such attacks, as they do not directly depend on parameters.
\vspace{-3mm}}
\label{fig:distrib_parameter}
\end{figure*}

We present the results of doing so in \Cref{fig:adaptive_samples} where two types of backdoor attacks and two levels of knowledge about the reference data points are considered. As one can see, the adversary may lower the distance between their backdoored and clean models by having a larger number of masked samples per reference data point. However, this also makes the optimization task harder since the adversary needs to force the backdoored model to resemble the clean surrogate with respect to more data, incurring a more significant computational cost.
However, regardless of whether the adversary uses a simple BadNet or a sophisticated Wasserstein backdoor, both attacks fail to make the distances between the backdoored model and the clean models close to distance$(W_c,W_c)$, demonstrating that these attacks struggle to bypass \name. Furthermore, we found that the adversary gains little from full knowledge of the reference data points used by Zest. As shown in \Cref{fig:adaptive_samples}, for both types of backdoors, having knowledge about the reference points does lead to smaller pairwise distances, but the difference is within one standard deviation.
We suspect this is a consequence of our use of `academic' datasets, in the sense that they are well curated (\eg they have few outliers, etc.) such that the reference data points and the points that the adversary guessed are from very similar distribution.

We also confirm that our findings are agnostic to the choice of model architecture and dataset. As in a VGG-11 on GTSRB (see above), the attack fails for a ResNet-20 on CIFAR-10--even when using 1000 samples per reference point, the distance$(W_b,W_c)$ does not move closer to distance$(W_c,W_c)$.

\subsection{Ablation Studies}
\label{ssec:eval-ablation}

In this section, we perform an ablation study of hyperparameters related to \name, including the distance metric, numbers of available servers, sub-runs lengths, training stages that the sub-runs start from, learning rates, and attack success rates. These studies explain the assumptions and design choices of \name, showing its applicability under a variety of conditions.

\looseness=-1
\para{Distance metrics}
We evaluate the performance of \name when integrated with 5 different distance metrics:
(1) cosine distance over model parameters, \ie parameter-space distance, (2) parameter-space distance after aligning the parameters with Git Re-Basin~\cite{ainsworth2023git}, (3) cosine distance over pre-softmax model predictions, \ie 
output-space distance, (4) the CKA similarity~\cite{kornblith2019cka}, and (5) Zest.

We start with (1) and (2), the cosine distance over (the Git Re-Basin transformation of) the model parameters. Git Re-Basin (2) transforms the parameters of a model to functionally equivalent parameters that lie in an approximately convex basin near another model. However, since models compared by \name are trained for a short sub-run from the same weight initialization, their parameters are already well-aligned. Empirically, we found that both parameter-space distances with or without alignment are nearly identical. We discover that for learning rates smaller than $0.01$ there are overlaps between distance$(W_c,W_c)$ and distance$(W_b,W_c)$, leading to errors in backdoor detection and even potential adaptive attacks, as shown in \Cref{fig:distrib_parameter}, making parameter-space distances poor choices for \name. In contrast, Zest does not suffer from this issue 
as it corresponds to model \textit{behaviors}.

\begin{figure}[t]
\centering
\includegraphics[width=0.95\linewidth]{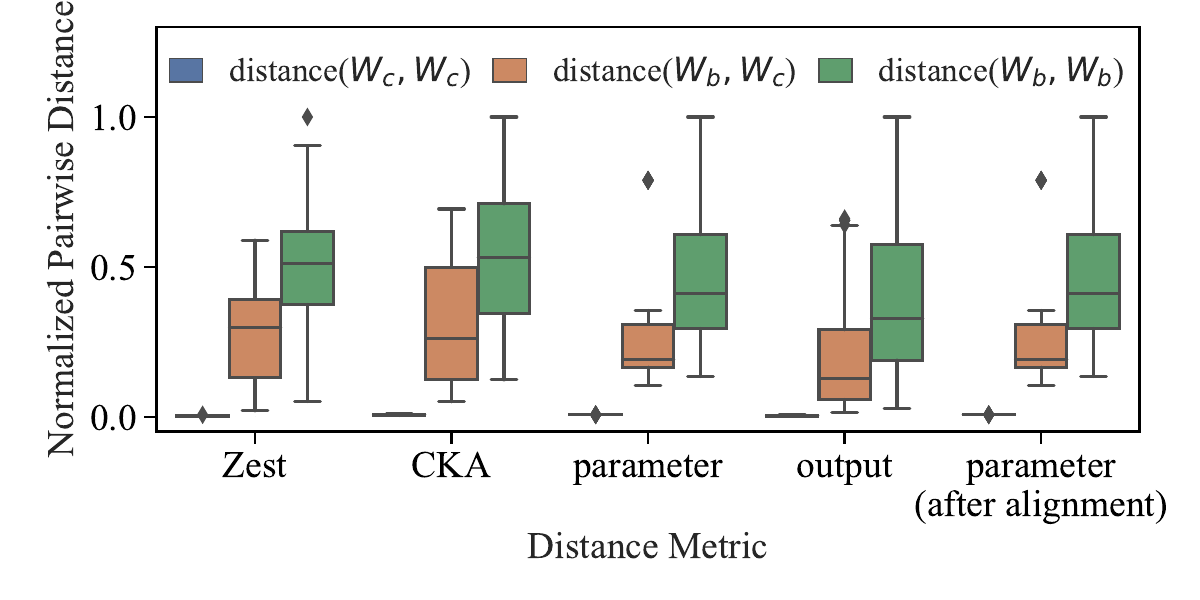}

\caption{
Boxplots of pairwise distances between the three groups (\ie namely, malicious distance$(W_b,W_b)$, benign-malicious distance$(W_b,W_c)$, and benign distance$(W_c,W_c)$) normalized to the range of 0 to 1. This figure simplifies \Cref{fig:main} but adds a dimension to allow direct comparison among different distance metrics.
It can be seen that all five metrics allow distance$(W_c, W_C)$ to form its own cluster which is separated from distance$(W_b,W_c)$, and distance$(W_b,W_b)$---this is a sign that \name will be able to correctly identify benign/malicious servers. However, to protect against adaptive adversaries, the Zest distance and CKA similarity should be preferred as the distance metric in \name. \vspace{-3mm}}
\label{fig:distances}
\end{figure}

When evaluating (3), the cosine output-space distance, we observe that the distance$(W_c,W_c)$ can form clusters that allow for detecting malicious servers, as shown in \Cref{fig:distances}.
However, compared to Zest, the output-space distance opens a larger attack space for adaptive adversaries. Consider the adaptive attack against Zest in~\Cref{ssec:adaptive_adv}; if the output-space and Zest distances were calculated over the same reference points, it would be more difficult for the adversary to fool Zest. This is because, in addition to the reference points, Zest would require that a pair of models behave similarly in the neighborhood of masked points around each reference point. 

While the CKA similarity can also form the cluster distance$(W_c,W_c)$ and detect non-adaptive adversaries, we did not identify an adaptive attack to evaluate it more thoroughly. However, CKA similarity is more computationally expensive and harder to implement than Zest as it needs the model's representation spaces whereas Zest only needs inference access. 

To summarize, we empirically demonstrated that parameter -space distances are ineffective compared to distances defined on outputs or representations (\ie output-space, Zest, and CKA), which are more suitable to \name. Furthermore, we show that Zest is resilient against the adaptive attacks in \Cref{ssec:adaptive_adv}, making it a suitable distance metric for \name.

\para{Number of Servers}
\looseness=-1
In Section~\ref{ssec:detection_eval}, we conduct experiments with 16 training servers, eight of which are benign. In practice, the client may reduce the number of servers to save costs or due to a limited supply of available servers. Therefore, we vary the number of servers and identify the proportion of benign servers needed for a detection accuracy exceeding $95\%$, which we show in ~\Cref{fig:servers}. Unsurprisingly, having fewer training servers requires a higher proportion of benign servers. It is worth noting that we find the absolute lowest number of required benign servers stays around six. To understand this, we consider a scenario with only five servers (see Figure~\ref{fig:5servers} in Appendix~\ref{app:additional_figures} for details). While our results from using five servers are consistent with our results in Section~\ref{ssec:detection_eval}, the low number of servers causes the KS-test to output higher significance values for both clean and backdoored models (\eg 0.2 for backdoored models and 1 for clean models). 

\looseness=-1
In such scenarios of limited servers, we replace the KS test with 
MAD outlier detection, which is commonly used in existing backdoor detectors. Specifically, we begin by identifying the minimum-variance cluster $c$ and the $r n - 1$ distances from the set $\{distance(W_{t+k,1}, W_{t+k,j}) |  j \in [2 \dots n] \}$ closest to it as usual. Then instead of performing a KS test, we compute the anomaly index for each distance from the set with respect to the benign cluster $c$ using MAD outlier detection. We then compare the third quartile of these anomaly indexes against the largest observed index of the values within the cluster to predict whether the primary server is benign. This leads to $100\%$ detection success, as shown in \Cref{fig:servers_mad} in Appendix~\ref{app:additional_figures}. However, like the subjective threshold choices in previous backdoor detectors, our choice of the third quartile is arbitrary and based on specific empirical results. Therefore, we cannot claim it would be a good choice across different settings, which is a disadvantage of MAD compared to the KS-test.

Lastly, we consider a special case where there are only three available servers and two are benign (\name would fail for a smaller number of servers or benign servers). In this case, there is only one observation each of distance$(W_c,W_c)$, distance$(W_b,W_c)$, and distance$(W_b,W_b)$, making it impossible to identify a cluster of distance$(W_c,W_c)$. A potential solution is to ``virtualize'' the servers, \ie submit multiple identical training tasks to each server. In other words, we relax the assumptions that servers do not collude and that malicious servers use different backdoors. By doing so \name, is able to detect the malicious server, though this may open new vulnerabilities. A more detailed discussion on the implications of relaxing this assumption can be found in \Cref{ssec:collude}.

\begin{table}[t]
\centering
\setlength{\tabcolsep}{3pt}
\resizebox{\linewidth}{!}{  %
\begin{tabular}{|c|cccccccc|}
\hline
\textbf{Number of Servers} & 8 & 10 & 12 & 14 & 16 & 18 & 20 & 22 \\ 
\hline
\textbf{\% Benign Servers} & 75.0 & 60.0 & 50.0 & 42.9 & 37.5 & 33.3 & 30.0 & 27.3 \\ 
\hline
\end{tabular}
}
\caption{\looseness=-1 
Percentage of benign servers needed to achieve detection accuracies above $95\%$ with respect to the number of training servers. There is a decreasing trend; at least six benign servers are required to maintain performance due to the sample-size properties of the statistical test used in \name.\vspace{-3mm} }
\label{fig:servers}
\end{table}

\looseness=-1
\para{Length of sub-run $k$}
Recall from Section~\ref{ssec:intuition} that 
the deviation of (clean) model updates caused by stochasticity in training may be limited by restricting the number of training steps, a.k.a., length of sub-run $k$. This hyperparameter allows the client to trade off the increased ability to detect malicious servers when the sub-runs are of smaller length (note that some servers may set a lower bound on the number of training steps and refuse shorter jobs) with the larger overhead of increased interactions with the server to submit the next sub-run request. We show in Figure~\ref{fig:ablation_k} in Appendix~\ref{subapp:subrun_len}
(see a more detailed discussion on setting $k$ in Appendix~\ref{subapp:subrun_len})
, that even with $k=4000$ (\ie around 10 epochs) for CIFAR10 where there are 200 epochs in total, distance$(W_c,W_c)$ still forms a distinct distribution from other pairwise distances.

\looseness=-1
\para{Stage of training (start $t$ of the sub-run)} Alongside the sub-run length, the stage that the sub-run occurs in also correlates with training stochasticity. At the start of training, models are randomly initialized and are likely to be near local loss maxima. Thus the gradients resulting from the first few minibatches are highly sensitive to the data they contain. In contrast, near the end of training, the model is ideally near a locally optimal state and is highly likely to keep approaching it. This indicates that the variance between updates decreases over the course of training, which agrees with our ablation of $t$, the training step the sub-run begins from. By testing \name at five stages of training, we observe larger variance for the benign cluster, distance$(W_c,W_c)$ when $t$ is small, \ie before the $50^{th}$ epoch (see Figure~\ref{fig:ablation_t} in Appendix~\ref{subapp:t}). This results in smaller p-values for some of the clean models during the KS test. However, we do not consider this a strong limitation, since (1) the detection is only wrong for one of the 32 models at the early stages of the training and (2) it is unlikely for adversaries to \textit{only} perform backdoor attacks at the very beginning of training. If they do, the backdoor may be removed due to phenomena such as catastrophic forgetting~\cite{french1999catastrophic} and natural forgetting~\cite{jagielski2022measuring}. We empirically verify that this is indeed the case on ResNet-20 with CIFAR-10.

\begin{figure}[t]
\centering

\includegraphics[width=0.95\linewidth]{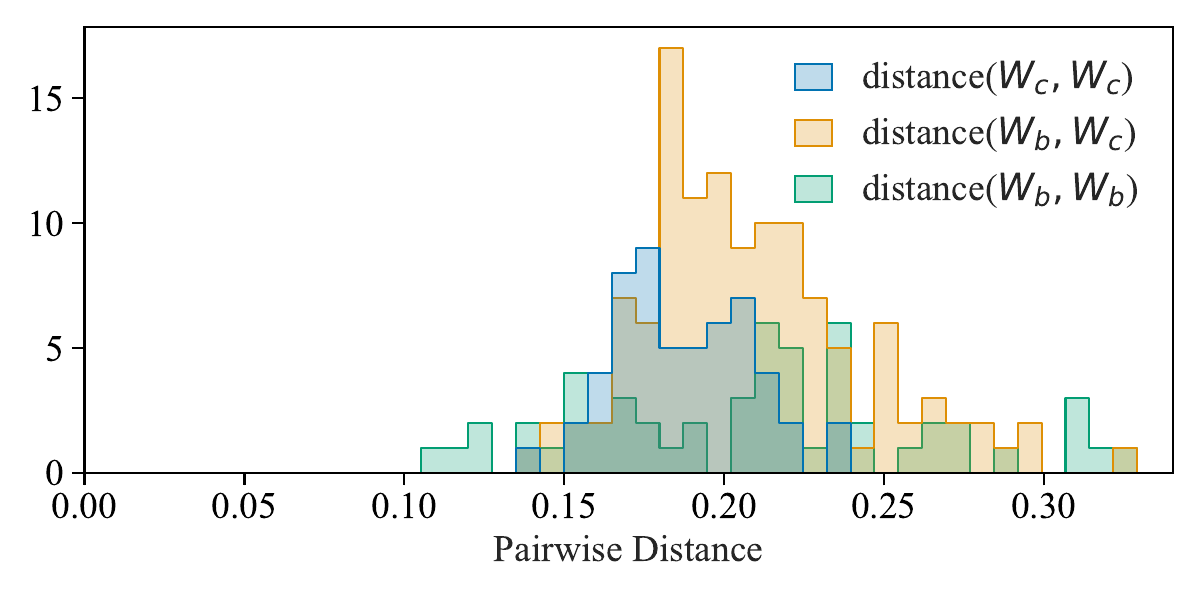}

\caption{\looseness=-1
Histogram of pairwise distances when the client increases $\eta$ by a factor of 10. It can be seen that, unlike in \Cref{fig:main}, the distances between clean models are large and \name is unable to approximate the benign cluster accurately, leading to performance degradation. Therefore, the client should carefully set $\eta$ as it may give malicious servers a large advantage.\vspace{-3mm}}
\label{fig:ablation_eta_client}
\end{figure}

\looseness=-1
\noindent\textbf{Learning rate $\eta$} The learning rate ($\eta$) can also impact model updates variance. Imagine a model with a locally convex loss landscape (\eg a model has already been trained for a while). A small learning rate $\eta$ will encourage updates toward the same local optima, whereas a larger $\eta$ may cause the model to ``jump'' from the locally convex area. Empirically, we increase the learning rate we use in earlier sections for CIFAR10 by a factor of 10 to $0.1$ and show the resulting pairwise distances in \Cref{fig:ablation_eta_client}. With larger learning rates, the distances among clean models vary far more, making it difficult to correctly identify the benign cluster of distance$(W_c,W_c)$. However, the learning rate used by benign servers is within the control of the client and a large $\eta$ is usually only used at the beginning of training--\eg setting $\eta=0.01$ instead of $\eta=0.1$ here allows the model to achieve better accuracy. Therefore, we consider this a minor limitation. 
We also study the impact of decreasing the learning rate and find it does not impact the cluster of distance$(W_c,W_c)$ significantly (more details can be found in Appendix~\ref{subapp:lr}). However, when malicious servers decrease the learning rate, they may move backdoored models' distances closer to distance$(W_c,W_c)$ for some distance metrics (but not Zest). This is studied as an adaptive attack in \Cref{ssec:adaptive_adv}.

\begin{figure}[t]
\centering

\includegraphics[width=0.95\linewidth]{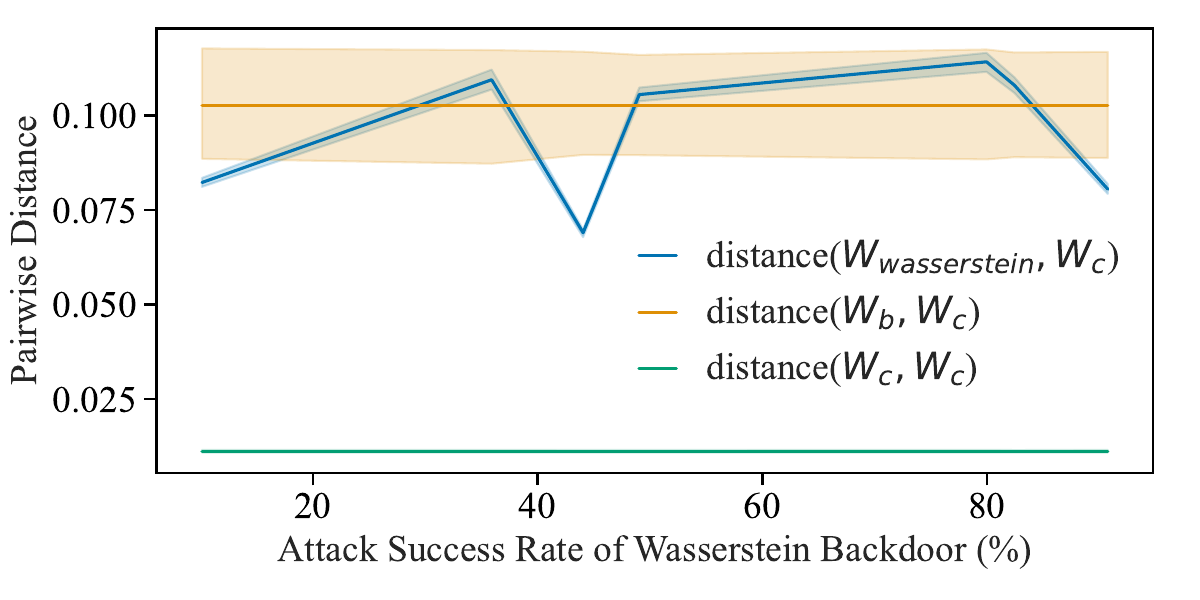}

\caption{The distance between Wasserstein-backdoored models and clean models as a function of the backdoored attack success rates (ASR). Average clean-backdoor and clean-clean distances are plotted as horizontal lines for reference. The distance$(W_{\textit{Wasserstein}},W_c)$ is not strongly correlated to the ASR and is consistently larger than distance$(W_c,W_c)$.\vspace{-3mm}}
\label{fig:asr}
\end{figure}

\looseness=-1
\noindent\textbf{Attack Success Rate} The attack success rate (ASR) of backdoored models can be manipulated by adversaries via hyperparameters associated with the attacks. One may expect a backdoored model with a smaller ASR to be closer to clean models, however, we observe the opposite for Wasserstein backdoors with varying ASRs, as shown in \Cref{fig:asr}. In this case, the distance from the Wasserstein-backdoored models to the clean models is always significant and independent of the ASR, even when the attack is no better than random guessing. This indicates that \name is robust to various ASRs.

\section{Discussion}
\label{sec:discussion}

\subsection{Relaxing the Assumption of Non-colluding Servers}
\label{ssec:collude}
\looseness=-1
Recall from \Cref{ssec:threat_model} that \name assumes the cloud providers do not collude. In particular, this implies that malicious servers do not use identical backdoors. Here we relax this assumption and consider the scenario that multiple returned models may be backdoored in the same way. Doing so is also helpful for \name when there are very few (\eg $n<5$) servers available.
For instance, in cases where $n=3$, the client may want to submit the same training job multiple times (\eg 5) to each server and to virtualize having more servers and run \name as if $n=15$. Naturally, if one of the three servers is malicious, it will backdoor the models using the same backdoor strategy, or even return the same backdoored model to all five training jobs it receives. 

\begin{figure}[t]
    \centering
    \subfloat[Pairwise Distance
    ]
    {
    \includegraphics[width=0.95\linewidth]{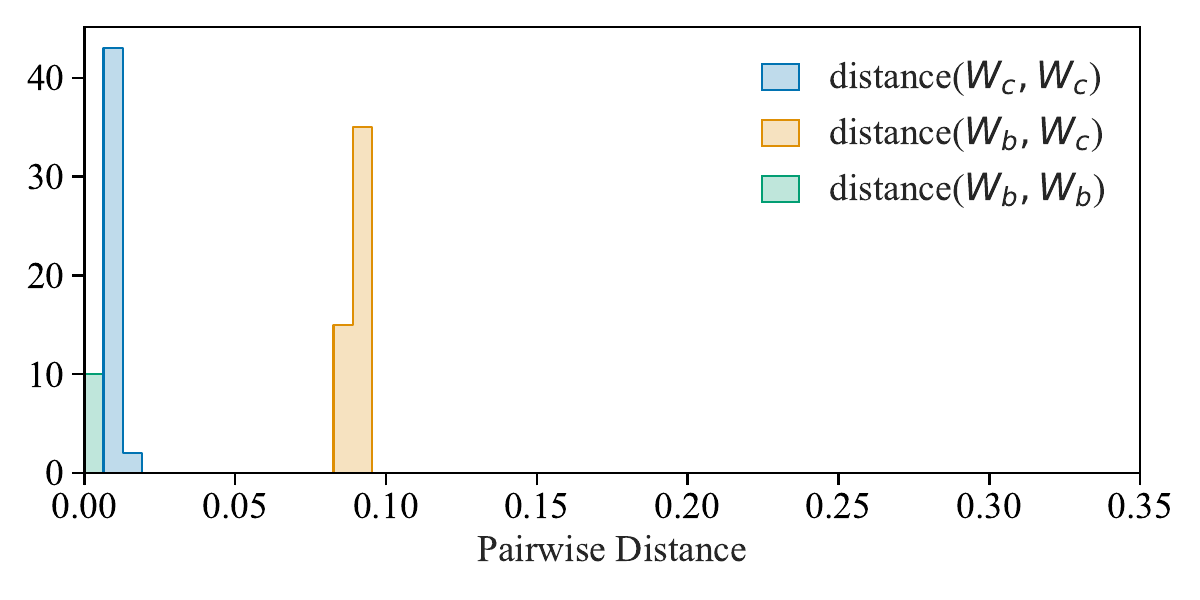}
    }
    
    \subfloat[Client's View \label{subfig1:cluster_3server}
    ]
    {
    \includegraphics[width=0.95\linewidth]{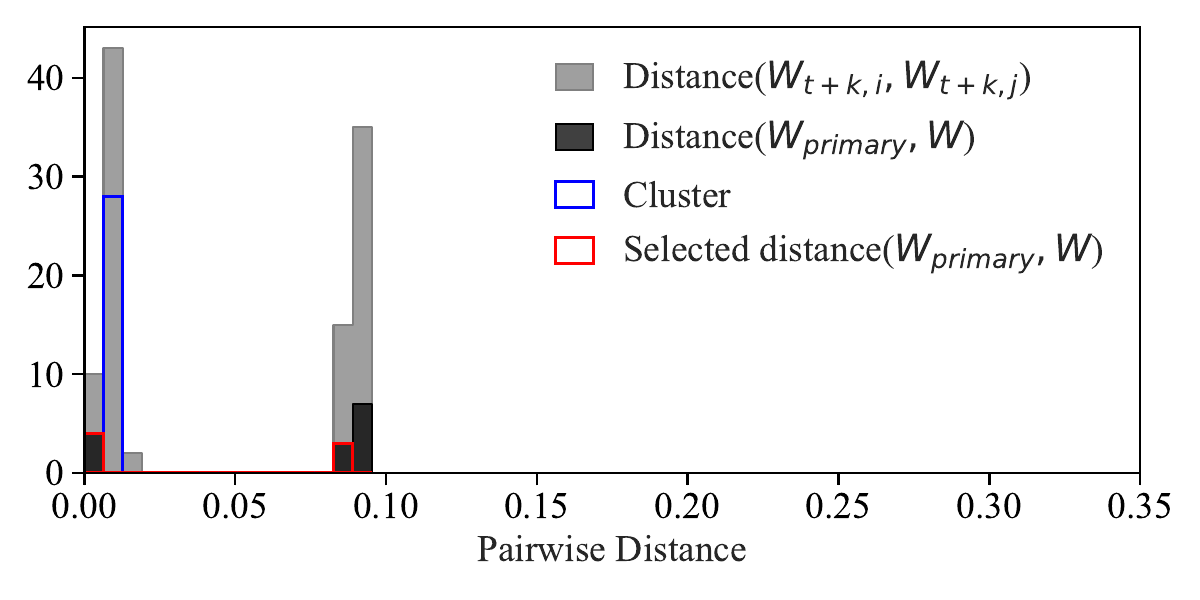}
    }
    \caption{
    Reproduction of Figure~\ref{fig:main} except there are only 3 available servers and the identical training task is repeated by each of them for 5 times. In such a scenario, the assumption of non-colluding cloud providers does not hold anymore, so we relax it and consider the scenario that the malicious server returns five identical backdoored models. Compared to Figure~\ref{subfig:main_benign}, here we observe significantly smaller variances for distance$(W_b,W_c)$ and distance$(W_b,W_b)$. However, \name is still able to detect this malicious server, as shown in Figure~\ref{subfig1:cluster_3server}.
    \vspace{-3mm}
    }
    \label{fig:3servers}
\end{figure}

\looseness=-1
Now we empirically evaluate \name in this scenario, or equivalently, the scenario that there exists $15$ servers and $5$ of them are malicious and collude with each other. \name is performed and the pairwise distances between the models returned is presented in \Cref{fig:3servers}. As one can see, the consequences of relaxing the non-collusion assumption are twofold:
(1) since the backdoor servers collude, distance$(W_b,W_b)$ is smaller than the distance between two models trained by non-colluding benign servers; and (2) distance$(W_b,W_c)$ also forms a cluster since all $W_b$ are from the same backdoored model. Since the majority of the servers are benign, \name can identify the cluster of distance$(W_c,W_c)$ and correctly detect the malicious server. Therefore, if the number of malicious servers is greater than or equal to the number of benign servers, there could be many incorrect detections. Compare this to the non-colluding case; if there are fifteen non-colluding servers, only $40\%$ of them need to be benign for high detection accuracy (see \Cref{fig:servers}). Relaxing this assumption may also lead to vulnerabilities that may be exploited by adaptive adversaries if they know the client would ``virtualize'' the servers. To conclude, \name is more robust when there are more non-colluding servers.

\subsection{Alternative Threat Model}

Perhaps one of the most important insights of our work is that we study the threat model where the backdooring adversary is a server to whom the client submitted training data. This departs from past work which considers dataset-level adversaries and which detect backdoors by analyzing the training data; in our threat model, the client does not know what data the servers used to train the outputted model or whether the server backdoored the model by changing the training algorithm. One of the key advantages of our threat model is that it allows our approach to operate without explicit knowledge of what distinguishes the backdoored from the legitimate distribution of model updates. This is a difficult open problem that shares many similarities with the challenges faced by intrusion detection systems~\cite{lee1998data} which attempt to distinguish anomalous from normal system behavior. 

Throughout this paper, we assumed \name would be deployed while the client outsources the training of a model. However, this threat model can be further expanded while still employing our method. For example, instead of detecting malicious servers as the model of interest is being trained, a client could aim to decide whether a given cloud provider could be trusted by challenging it prior to using its compute services. This could be done by employing \name on a (possibly synthetic) training set. This initial training run would only be used to infer whether the server can be trusted or not. Once each cloud provider has been evaluated, the client can decide to only use the servers from cloud providers the client trusts. This of course assumes that a cloud provider that was initially benign does not later become adversarial.

\subsection{Verified Computing}
\looseness=-1
Identifying malicious servers in our threat model consists of detecting that they deviated from the computations expected of them. That is, our approach is designed to ascertain that a server did not insert a backdoor because it: 1) used the training data submitted to it, and 2) used the training algorithm faithfully on this data (computed the updates correctly, followed expected randomness, etc.). These are problems often tackled by verified computing, which is a known solution for other problems in trustworthy machine learning such as preserving data confidentiality~\cite{juvekar2018gazelle,kim2018logistic}. An alternative approach to backdoor detection could also require the cloud providers to use verified computing as they run the training algorithm, providing guarantees that the servers executed training faithfully. However, the methods used to verify the computation of deep learning training algorithms are currently prohibitively expensive.

\section{Conclusions}
\looseness=-1
We considered a novel threat model for backdoor detection where the adversary is a server that provides ML as a service. Clients who seek to outsource the training of a model to an untrusted, potentially malicious server can leverage \name to identify any malicious servers they interact with. Intuitively, our approach is most effective for stages of training where model updates start to converge (\eg later on in training and when the learning rate has decreased). We find that despite making few assumptions about the training algorithm, \name is able to successfully detect malicious servers across different training settings in both vision and language domains as long as there are $\geq 3$ servers while not all of them are malicious. We believe future work will be able to demonstrate the versatility of the approach we propose here by adapting \name to other paradigms of learning, such as self-supervised learning. 

\section{Acknowledgements}
We would like to acknowledge our sponsors, who support our research with financial and in-kind contributions: CIFAR (through the Canada CIFAR AI Chair), Ericsson, NSERC (under the Discovery Program), Ontario (through the Early Researcher Award), and the Sloan Foundation. Resources used in preparing this research were provided, in part, by the Province of Ontario, the Government of Canada through CIFAR, and companies sponsoring the Vector Institute.

\bibliographystyle{IEEEtranS}
\bibliography{main}

\appendices
\newpage
\section{Experimental Platform.}\label{ssec:exp_platform}
Experiments were performed with Ubuntu 18.04.6 LTS using 16 Intel Xeon CPU cores and 4 NVIDIA T4 GPUs with 16 gigabytes of memory each. Our codebase was implemented with Python 3.9 and PyTorch, using the LIME library~\cite{RibeiroLIME} to support the Zest distance. 

\section{Model Distance Metrics}
\label{app:model_distances}
A key aspect of our approach relies on its ability to compare the models returned by different cloud providers. 
To this effect, we review the existing literature on computing distances between models to quantify similarities in models' predictive behavior~\cite{xie2019diffchaser, pei2017deepxplore,li2021modeldiff,jia2021proof}. 

There are several challenges involved with comparing models directly. First, because we may be interested in comparing models whose architectures are different (\eg two DNNs with different numbers of hidden layers), we cannot directly compare model parameter vectors. In fact, model parameter distances can be meaningless even when the models share the same architecture~\cite{jia2021proof}. Take the example of a DNN where the neurons on one of its hidden layers are permuted to yield a second DNN with a very different parameter vector and a large distance in the parameter space between the DNNs. Yet, the two models will produce identical outputs on all inputs they are given: the large parameter distance does not accurately reflect their identical behavior. To avoid this, it may be tempting to compare models by their predictions on the test distribution. The test distribution must necessarily be approximated by sampling --- making prediction comparison difficult due to high variance and high dependence on the representativeness of the sampling. 

Another notion of model distance known as Zest~\cite{jia2022a} leverages a popular model explanation technique, local interpretable model-agnostic explanations (LIME)~\cite{RibeiroLIME}, to circumvent these challenges. This technique interpolates between parameter and prediction space comparisons by creating local approximations of the models to compare their global behavior. 
To approximate the global behavior of a model, a set of data points is sampled from the training distribution, and a linear model is trained for each of these data points. Specifically, Zest segments the data point (which is assumed to be an image), randomly masks out some of the segments, and queries the ML models to be compared for labels on these masked data points. Then, a linear model is trained using binary vectors indicating which segment is masked out as inputs and labels given by the model as outputs. Intuitively, each element of the weight matrices of this linear approximation indicates how a segment of the data point contributes to a certain class in the model. This returns many linear models, each capturing the behavior of the model in a different part of its input domain. The weight vectors of these linear models can be concatenated together to form an approximation of the global behavior of the model---or a signature of the model. To compare the predictive behavior of models, a cosine distance is computed between their signatures. This focus on functionality is the main reason why we choose to use it in this work: since backdoored models are trained to learn a separate and potentially unique backdoor task in addition to the primary training task, leveraging Zest facilitates the detection of malicious servers. 
\section{Additional Ablation Studies}
\label{app:additional_ablation}

\subsection{Length of sub-run $k$}
\label{subapp:subrun_len}
\begin{figure}[t]
\centering
\includegraphics[width=0.95\linewidth]{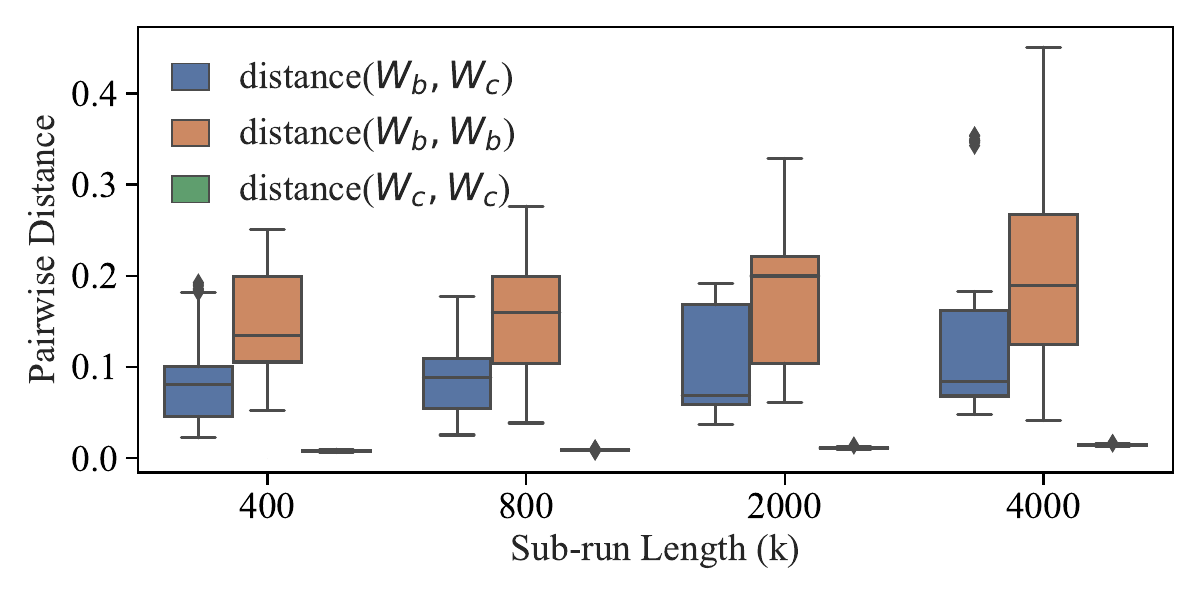}

\caption{Boxplot of pairwise distances: namely, malicious distance$(W_b,W_b)$, benign-malicious distance$(W_b,W_c)$, and benign distance$(W_c,W_c)$. These are computed with varying lengths $k$ of the sub-run, instead of fixing it to 2000 steps as done in \Cref{ssec:detection_eval}. The rest of experimental setup is identical. As can be seen, the distributions of distance$(W_c,W_c)$ do not vary significantly as  $k$ increases from 400 to 4000, whereas the variances of distance$(W_b,W_b)$ and distance$(W_b,W_c)$ increase slightly.}
\label{fig:ablation_k}
\end{figure}

Here we study the impacts of the length of the sub-runs, $k$, in terms of the detection performance of \name and its cost to the client. Recall that \name relies on the small deviation among clean models as caused by randomness in training, which is correlated to the number of training steps. Intuitively, if the client sets $k=1$, many sources of randomness in training such as data ordering may be avoided. However, this is not realistic since (1) this would incur an extremely high communication cost between the client and servers, and (2) some servers may not accept such a short training job. Therefore, in this work, we consider $k$ to be a multiple of the number of training steps in one epoch, which is about 400 steps for CIFAR-10 when the mini-batch size is $128$. In \Cref{fig:ablation_k}, we plot the distributions of the pairwise distances with $k \in [400, 800, 2000, 4000]$, where it can be seen that the small variance in distance$(W_c,W_c)$ consistently holds with respect to $k$. Therefore, if the client aims to reduce the overhead of communicating with the servers, a larger $k$ is preferable.

\subsection{Stage of training (start $t$ of the sub-run)}
\label{subapp:t}
In \Cref{ssec:detection_eval}, we evaluate \name on model checkpoints obtained after 100 epochs of training. To understand the impact of the stage of training from which the sub-run is selected, we vary the number of epochs ($t$) the models were trained for prior to detection. After each server performs a sub-run on these models, we compute the pairwise distances among them and analyze \name's effectiveness.

\begin{figure}[t]
\centering
\includegraphics[width=0.95\linewidth]{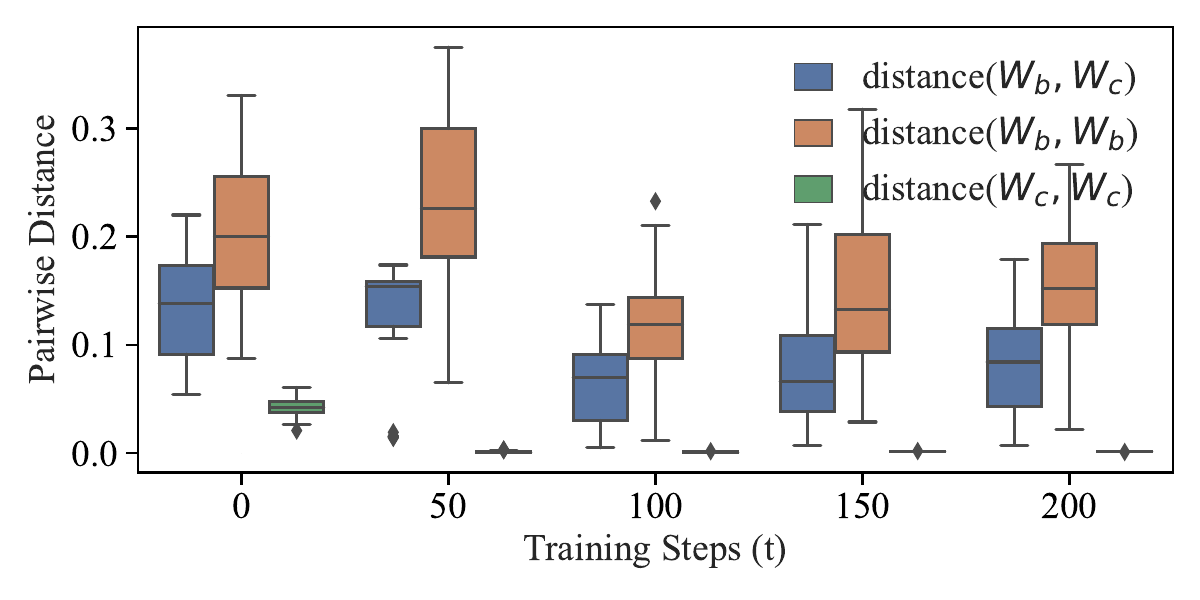}

\caption{ For $t>=50$, the variance of distance$(W_c,W_c)$ is consistently small which enables \name to function well. However, at the beginning of training when the model is at a random state(\ie $t=0$), the variance of distance$(W_c,W_c)$ is larger, meaning our method may  not perform as well early on in training.}
\label{fig:ablation_t}
\end{figure}

These pairwise distances are plotted as a box plot in \Cref{fig:ablation_t}. The distributions of the three categories of pairwise distances are consistent with those presented earlier at $t=100$. This is true for all stages of training except at the beginning of training when $t=0$ and the model is randomly initialized. At this stage of training we observe a large variance in the benign cluster (distance$(W_c,W_c)$). This is likely due to the varying influence of noise during training. At later stages in training, model weights may be close to a local optimum and continuing their training will likely push them closer to this optimum despite the stochasticity of training. However, at early stages of training the gradient descent updates computed for a randomly initialized model depend strongly upon randomness in the training procedure. For example, the choice of points to form the first minibatch of data may significantly influence the model to move in distinct directions in its parameter space if there are multiple local optima around the initialization. Similar observations were found in the paper introducing the Zest distance~\cite{jia2022a}: their Figure 2 shows a rapid change in model distance at its first few epochs of training. 

Empirical results also confirm that such a large variance of distance$(W_c,W_c)$ makes detecting malicious servers harder. In \Cref{ssec:detection_eval} where $t=100$ we get a detection accuracy of $99.6\%$, yet we can only achieve an accuracy of $74.0\%$ when $t=0$. This is not necessarily a strong limitation. If a malicious server manages to evade detection and insert a backdoor at the beginning of training, this backdoor is likely not robust to the future training updates required to achieve good performance on the model's task. Phenomena like catastrophic forgetting~\cite{french1999catastrophic} and natural forgetting~\cite{jagielski2022measuring} will make backdoors successfully inserted early on in training difficult to preserve without reinforcing them at later stages of training --- at which point the malicious server will no longer be able to bypass detection as it deviates from the agreed upon training algorithm.

\subsection{Learning rate $\eta$}
\label{subapp:lr}
At a given point in training, the learning rate $\eta$ also impacts the amount of variance between models trained by different servers from the same weight initialization. 
We therefore need to understand how the increased variance in models trained using larger learning rates impacts our ability to distinguish between the pairwise distance clusters.

\begin{figure}[t]
\centering
\subfloat[Client varies learning rate
]
{
\includegraphics[width=0.95\linewidth]{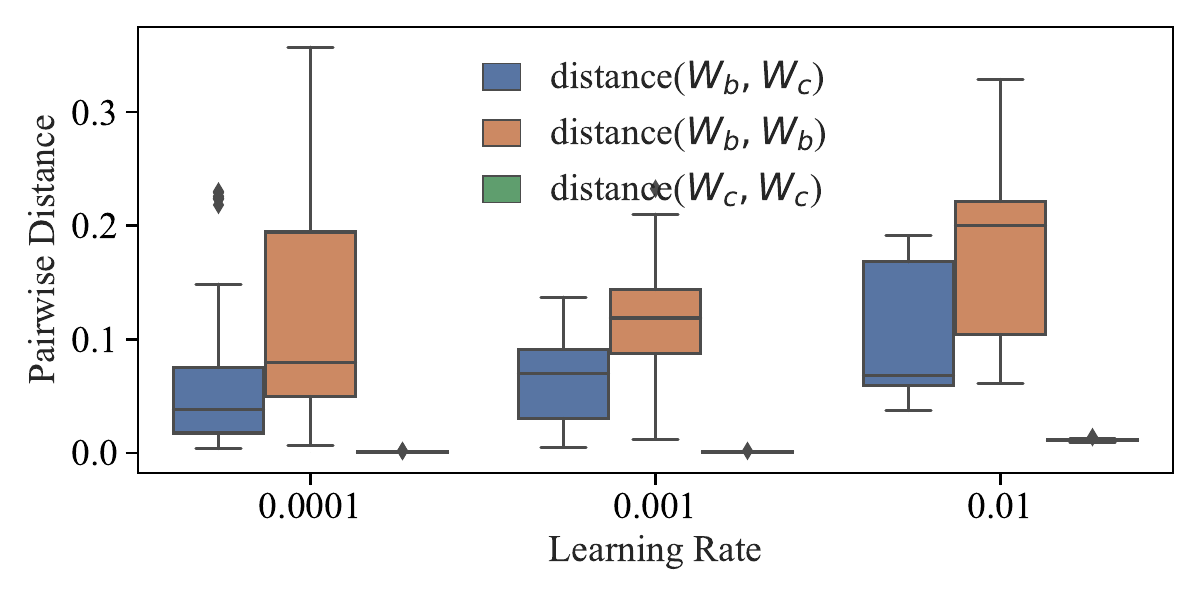}
}

\subfloat[Adversary varies learning rate
]
{
\includegraphics[width=0.95\linewidth]{figures/ablation_advlr_resnet20_CIFAR10_zest.pdf}
}
\caption{Boxplot of pairwise distances: namely, malicious distance$(W_b,W_b)$, benign-malicious distance$(W_b,W_c)$, and benign distance$(W_c,W_c)$. These are computed for malicious servers that train the backdoored models with different learning rates $\eta \in [0.01, 0.001, 0.0001]$, whereas the clean models are always trained with $\eta=0.01$. There is no significant difference between the distributions of pairwise distances, where one of the models is backdoored, trained with $\eta=0.01$ or $\eta=0.001$. Decreasing the learning rate further to 0.0001 reduces the variances of both distance$(W_b,W_b)$ and distance$(W_b,W_c)$, however, the variances are still larger than the variance of distance$(W_c,W_c)$.}
\label{fig:ablation_eta}
\end{figure}

In Section~\ref{ssec:detection_eval}, our experiments set the learning rate at $0.01$. Intuitively, reducing the learning rate should further reduce the distance$(W_c,W_c)$. We instead study the model distances when the learning rate is increased by a factor of 10 to $0.1$, which is shown in \Cref{fig:ablation_eta_client}. As expected, the distances among clean models now vary far more, making it difficult to correctly approximate the benign server cluster of distance$(W_c,W_c)$. Consequently, the detection is more likely to be inaccurate in settings where the learning rate is set to a large value. 

However, the learning rate is within the control of the client as they set the hyperparameters of the training algorithm. Large learning rates are typical in early stages of training and it is a common practice to gradually decrease the value of the learning rate as training progresses. Hence, as stated before when evaluating the effectiveness of our approach at different stages of training, we believe the impact of large learning rates on the effectiveness of our approach is a minor limitation: backdoors that are inserted earlier on in training are less likely to be robust given catastrophic forgetting. 

Furthermore, if a malicious server changes the learning rate their outputted model will differ from the benign servers that respected the client's learning rate. 
We evaluate this setting for a malicious server that selects a \textit{smaller} learning rate than the client. 
This is not only a more difficult detection challenge than an adversary using an increased learning rate; it is also a choice that can allow the malicious server to learn a backdoor through techniques adapted from transfer learning. 
In transfer learning, a small learning rate enables adapting to the new task without significantly modifying the model.
When translated to the setting of backdoor attacks, this small learning rate means that a backdoor can be easily inserted without degrading the model's performance on test data. 
We simulate a malicious server that uses equal or smaller learning rates than the one specified by the client. 
We plot the results in \Cref{fig:ablation_eta}. 
We find that reducing the learning rate to $0.0001$ leads to smaller variances for distance$(W_b,W_b)$ and distance$(W_b,W_c)$. 
However, this does not enable the malicious servers to bypass detection as the variance of distance$(W_c,W_c)$ remains the smallest. One explanation could be that for a model to learn a backdoor task, it has to move from a local optimal corresponding only to the classification task to another one that is locally optimal for both tasks (and not likely to be reached by benign training), no matter how smaller the learning rates are. %

\section{\name is General: a Case Study with Inspection-Based Backdoor Defenses}\label{ssec:ed_outsourced}
In this section we outline an alternative to the model distance metrics and KS-test used in \name (see Section~\ref{ssec:distribution}). Note that we find in our experiments that the Zest and CKA instantiations of our framework outperform the following proposals -- we only present the below for the sake of completeness. 

We propose to use the features of existing defenses which can characterize the triggers of some backdoor attacks and can be coupled with MAD outlier detection to identify malicious servers. As described in~\Cref{ssec:threat_model}, Neural Cleanse~\cite{Wang2019Neural}, TABOR~\cite{guo2020tabor}, and Neuron Inspect~\cite{huang2019neuroninspect} are relevant to our threat model as they only require access to the model and some clean testing data. These inspection-based backdoor defenses identify a plausible backdoor trigger for each class, measure the triggers according to a trigger feature, then perform anomaly detection to identify which trigger and class were used to backdoor the model, if any. 

As noted earlier, these defenses are evaded by several more recent attacks that use triggers that are implausible according to the assumptions the defenses make on backdoor triggers; we show in~\Cref{ssec:existing_defenses} that they can be fooled if a training-level adversary lowers the backdoor learning rate to weaken the attack signal. Regardless, we demonstrate the generality of our \name framework by adapting these existing defenses to our framework.  %

The defenses cannot be applied without some changes. As we are only interested in knowing whether the model is compromised, we can take the average over the classes of each model's trigger feature. This provides us a distribution over models of the average trigger feature. Using the average is an intentional choice; we seek to leverage an empirical mean estimator's lack of robustness to outliers. The average trigger feature of a model containing backdoored classes with outlier trigger features will differ substantially from the average trigger feature of the non-backdoored classes. In fact, this instability is why existing defenses prefer to measure outliers from an interval around the median rather than mean. 

With the distribution over the average trigger features, we use MAD to identify outliers (\ie models that present a backdoored class). Consistently with the literature, we use an anomaly index threshold of 2, corresponding to the 95\% confidence interval of the normally distributed feature data. 

This alternate method has several limitations not present in the version of \name proposed so far. Firstly, it will only be able to detect the attacks that the defense can identify normally; in the case of a training-level adversary, this method would be fooled if the adversary lowers the backdoor learning rate. Secondly, the number of malicious servers should be smaller than the number of benign servers so that MAD can still detect the outliers. Computationally, the largest cost is of identifying the potential triggers for each class. While Neural Cleanse and TABOR did not provide cost analysis, it has been shown that Neuron Inspect is about 10 times faster than them~\cite{huang2019neuroninspect}. The saliency map computation in Neuron Inspect requires one forward and backward pass for each data point, which is equivalent to the cost of one training step. Therefore, the cost of running this procedure is at least $D \cdot n \cdot m$ training steps, where $D$ is the number of mini-batches used.

\begin{figure}
    \centering
    \includegraphics[width=0.95\linewidth]{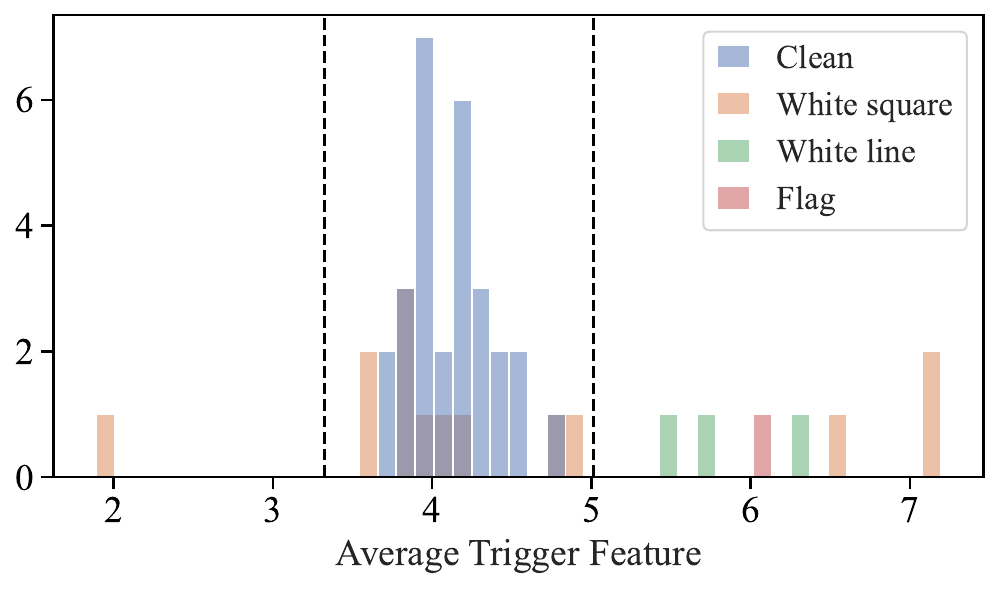}
    \caption{The distribution over the average trigger feature values for several clean and backdoored models. The black, dashed lines show the an interval extending a distance of $2\cdot$MAD from the median in either direction. Any point within this interval would be considered benign. The backdoored models that fooled this method were trained with lower backdoor learning rates. This result uses Neural Cleanse trigger feature, the $l_1$ norm, on CIFAR10 on 30 benign and 15 backdoored servers.}
    \label{fig:our_method_with_existing_defense}
\end{figure}

\looseness=-1
We next evaluate the alternate version of \name. With each returned model update, the client can run a backdoor defense mechanism to gather a distribution over the models of the average trigger feature value. Afterwards, the client can use MAD to identify the outlying models. As we can see in~\Cref{fig:our_method_with_existing_defense}, this method is only partially successful. As mentioned earlier, this method is limited by the sensitivity of the trigger figure to backdoor triggers. For example, Figure 5 of the paper that proposed the Wasserstein backdoor attack shows that their backdoored model had an anomaly index similar to or even lower than the clean models~\cite{khoaWasserstein}. 
In~\Cref{fig:our_method_with_existing_defense}, the thresholds on the trigger feature (the black, dashed lines) arising from $2\cdot$MAD can correctly identify several attacks. However, the method is fooled by attacks that used a lower backdoor learning rate, which overlap with the benign distribution. This is because the trigger feature value for these classes was close enough to the average trigger feature for the non-backdoored classes that the average over all the values did not significantly differ from the average trigger feature for the non-backdoored classes.

\section{Additional Figures}
\label{app:additional_figures}

\begin{figure}[h!]
    \centering
    \includegraphics[width=0.95\linewidth]{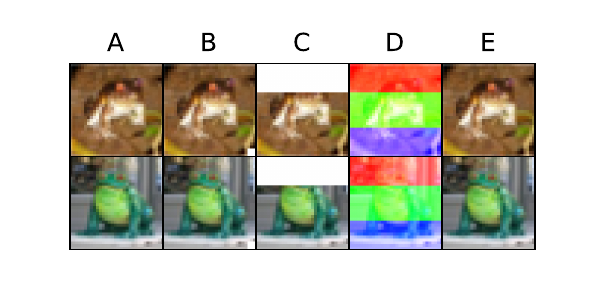}
    \caption{Original data alongside four backdoor triggers. a) clean image, b) white square BadNets attack, c) white stripe BadNets attack, d) RGB flag BadNets attack, and e) Wasserstein backdoor (expected to look like the clean image).}
    \label{fig:trigger}
\end{figure}

\begin{figure}[h!]
\centering
\includegraphics[width=0.95\linewidth]{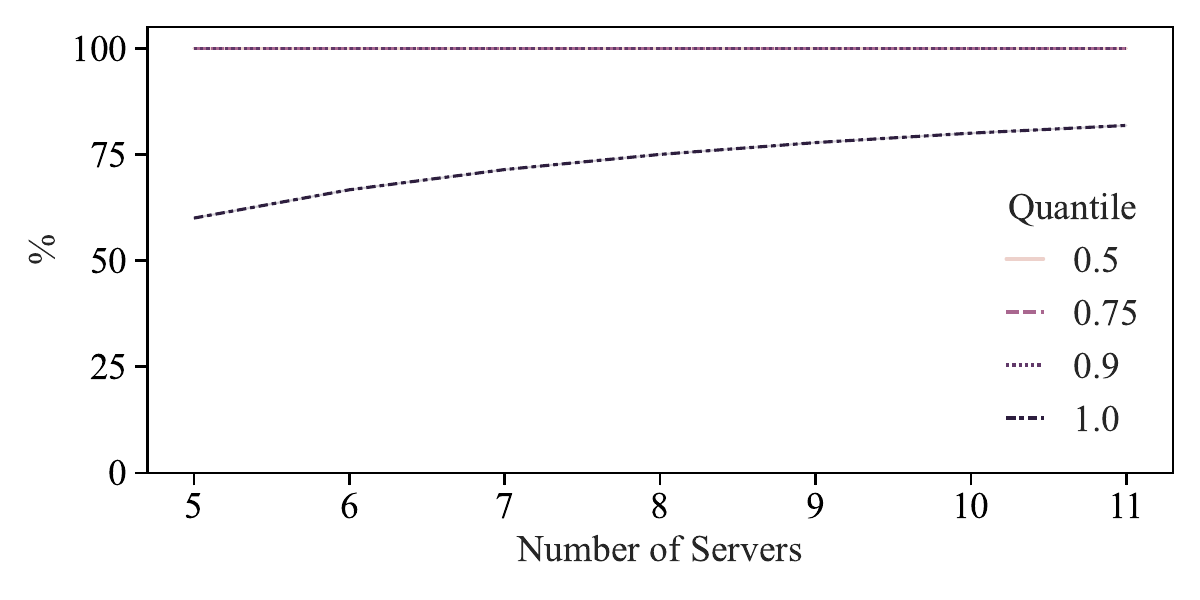}
\caption{Accuracy of backdoor detection when replacing the KS-test in \name with MAD outlier detection. The number of servers the client has access to is plotted on the x-axis, where at least are benign. If we predict that a model is backdoored when the third quantile of its anomaly indexes is above the threshold, the detection can be perfectly accurate as long as there are at least five servers.}
\label{fig:servers_mad}
\end{figure}

\begin{figure}[h!]
    \centering
    \begin{subfigure}[b]{0.95\linewidth}
       \includegraphics[width=1\linewidth]{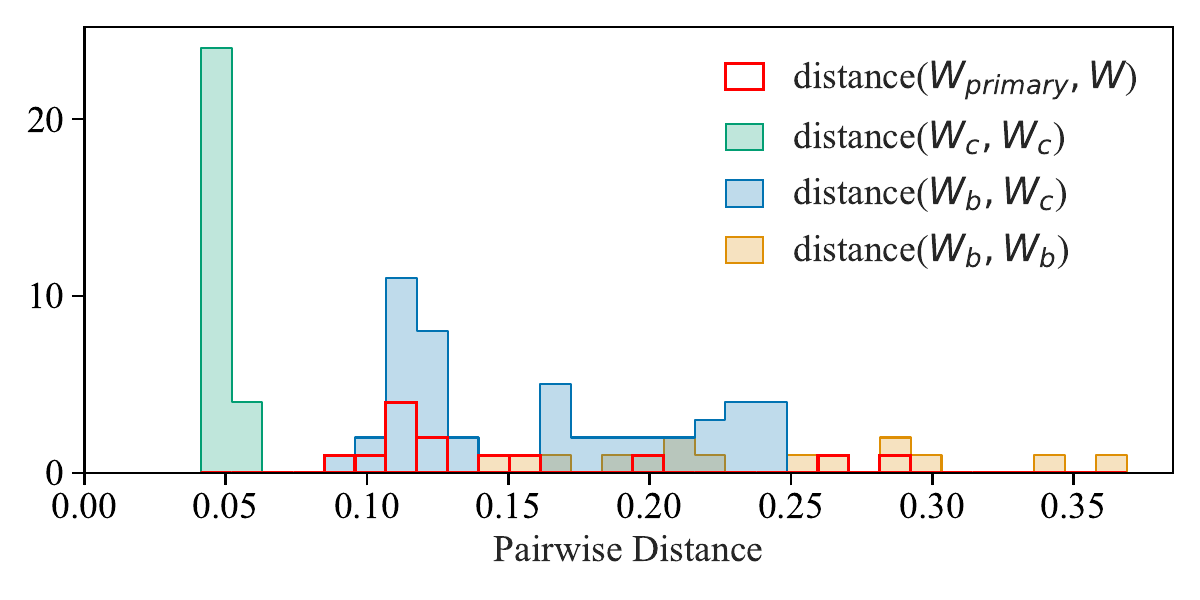}
       \caption{Pairwise Distance}
       \label{subfig:main_malicious} 
    \end{subfigure}
    \begin{subfigure}[b]{0.95\linewidth}
       \includegraphics[width=1\linewidth]{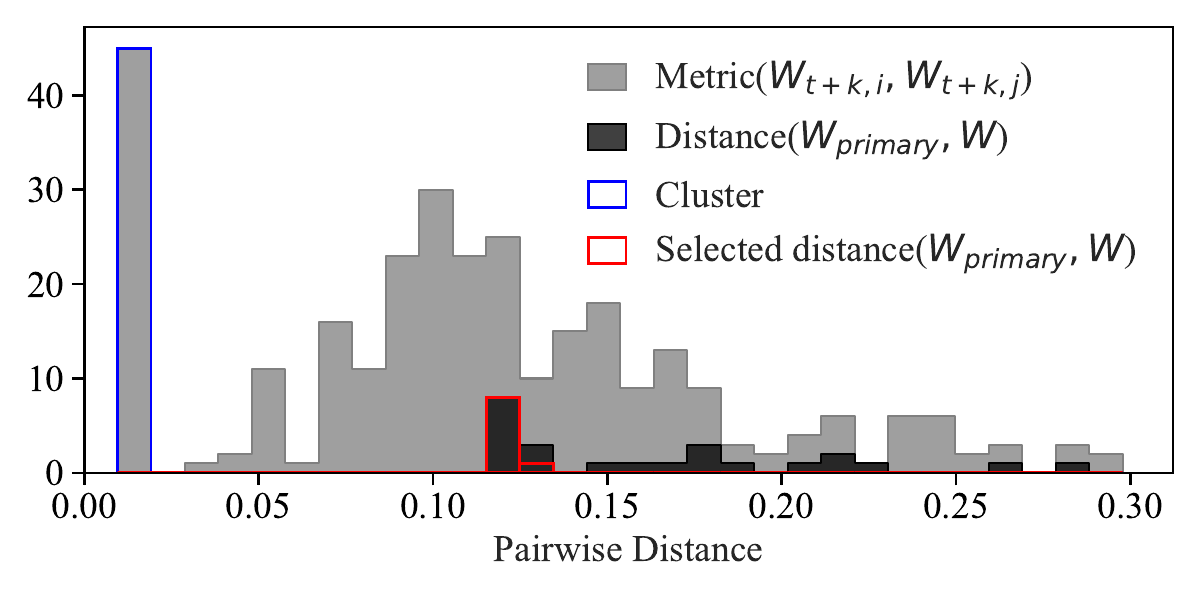}
       \caption{Client's View}
       \label{subfig1:cluster_malicious} 
    \end{subfigure}
    \caption{
    Reproduction of \Cref{fig:main} except the primary server is malicious.
}
\label{fig:main_malicious}
\end{figure}

\begin{figure}[h!]
\centering
\subfloat[Benign Primary Server
]
{
\includegraphics[width=0.95\linewidth]{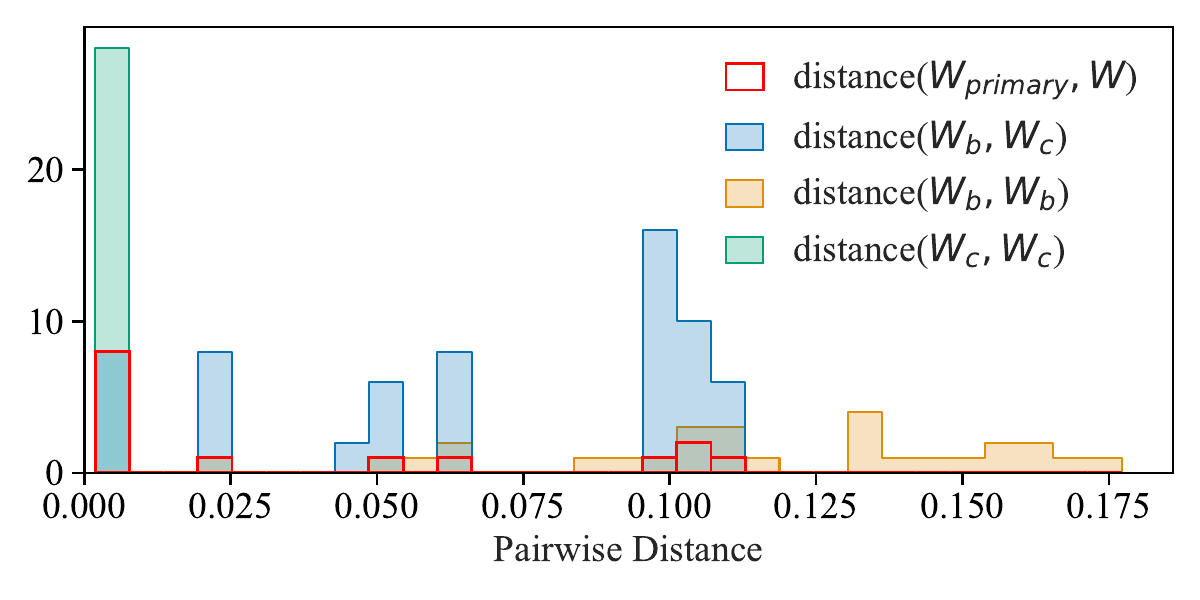}
}

\subfloat[Malicious Primary Server
]
{
\includegraphics[width=0.95\linewidth]{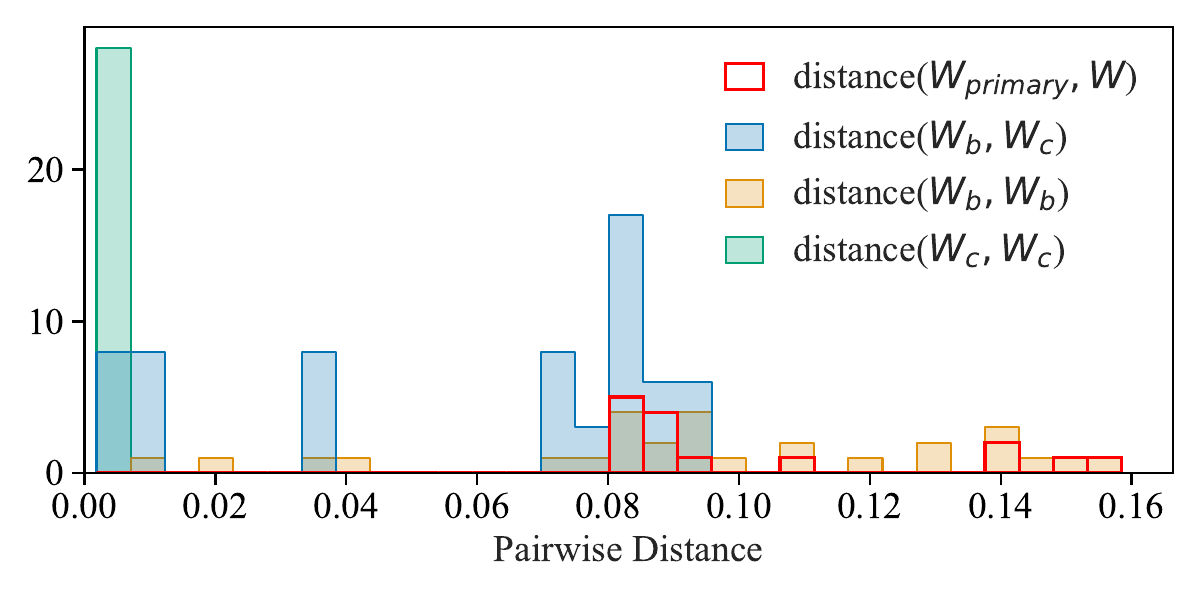}
}
\caption{Reproduction of Figure~\ref{subfig:main_benign} on VGG-11 models trained on the GTSRB dataset for both benign and malicious primary servers. The results are consistent.}
\label{fig:main_gtsrb}
\end{figure}

\begin{figure}[h!]
\centering
\subfloat[Benign Primary Server
]
{
\includegraphics[width=0.95\linewidth]{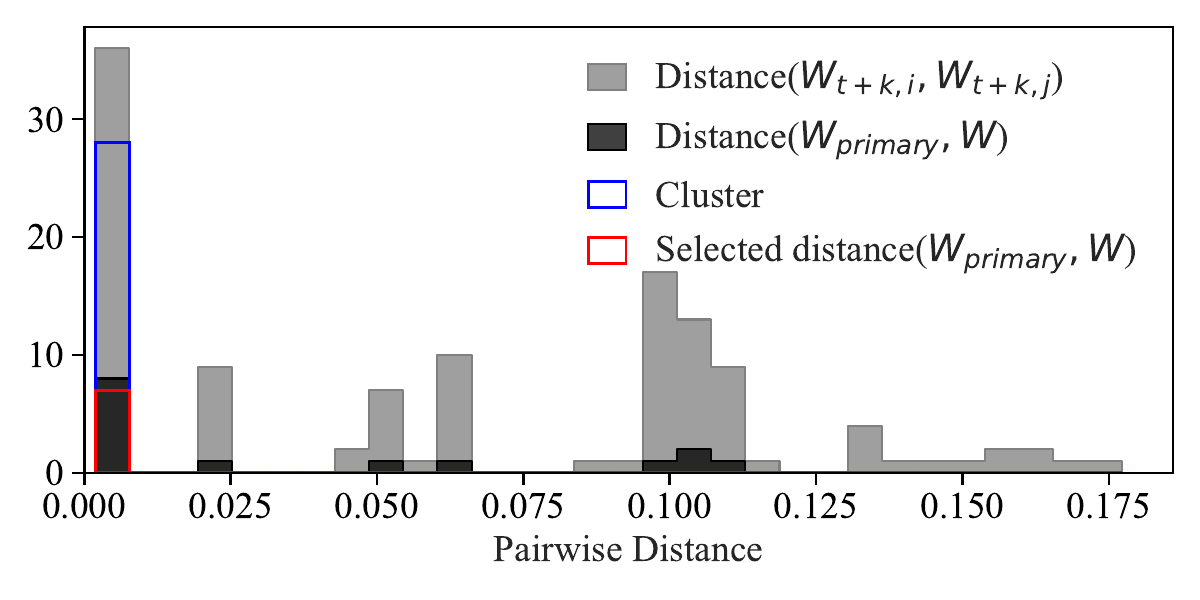}
}

\subfloat[Malicious Primary Server
]
{
\includegraphics[width=0.95\linewidth]{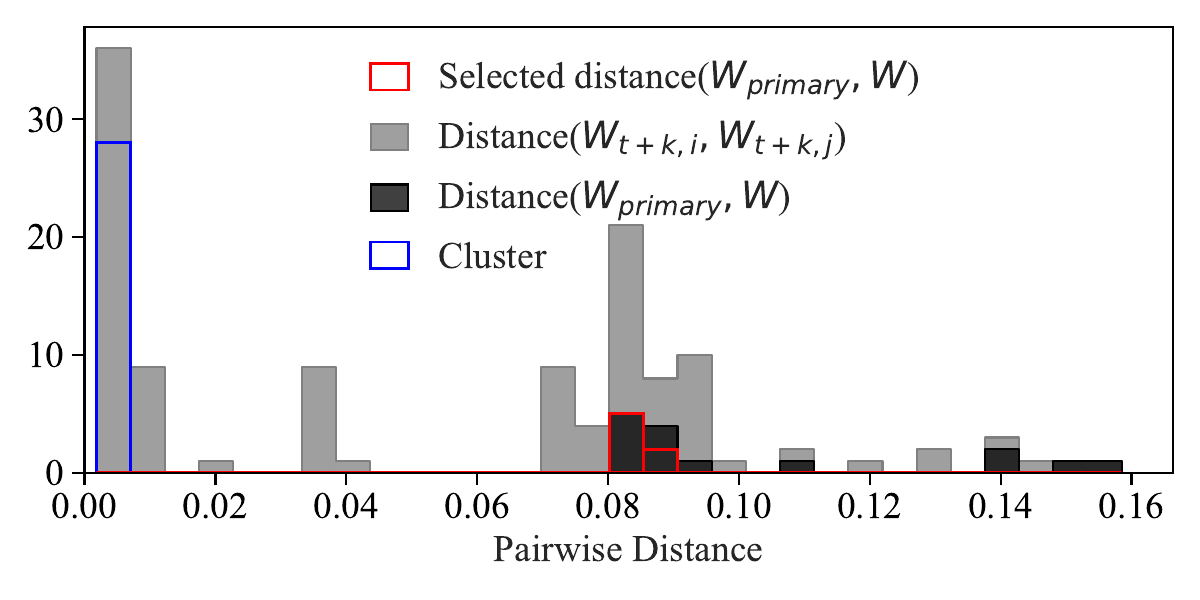}
}
\caption{Reproduction of Figure~\ref{subfig1:cluster_benign} on VGG-11 models trained on the GTSRB dataset from the client's point of view for both benign and malicious primary servers. The results are consistent.}
\label{fig:cluster_gtsrb}
\end{figure}

\begin{figure}[h!]
\centering
\subfloat[Pairwise Distance
]
{
\includegraphics[width=0.95\linewidth]
{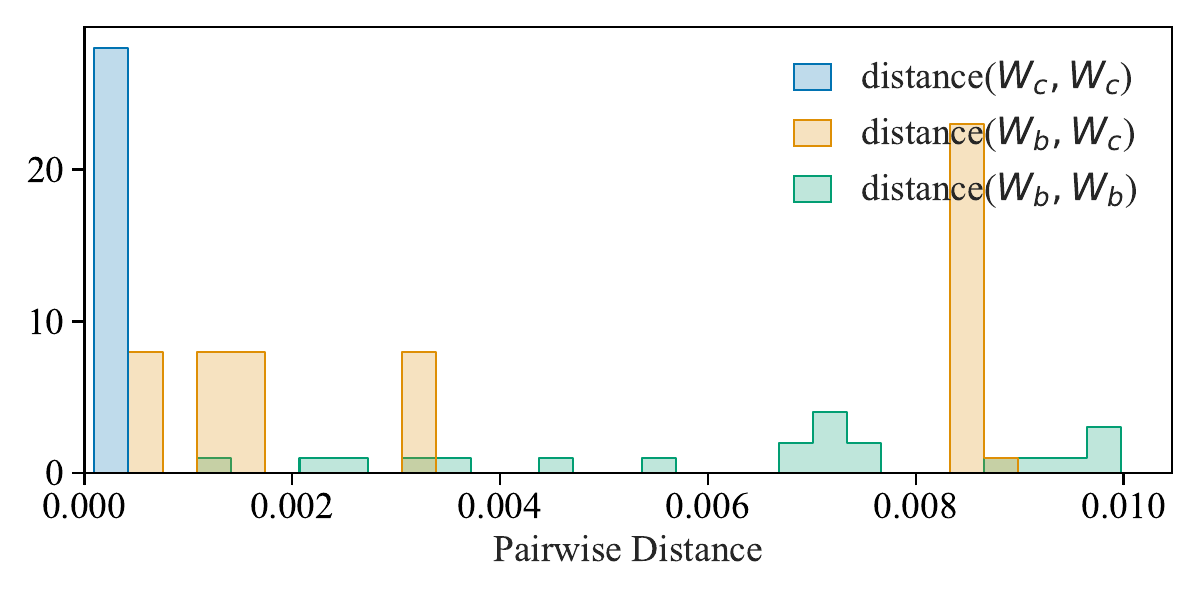}
}

\subfloat[Client's View
]
{
\includegraphics[width=0.95\linewidth]
{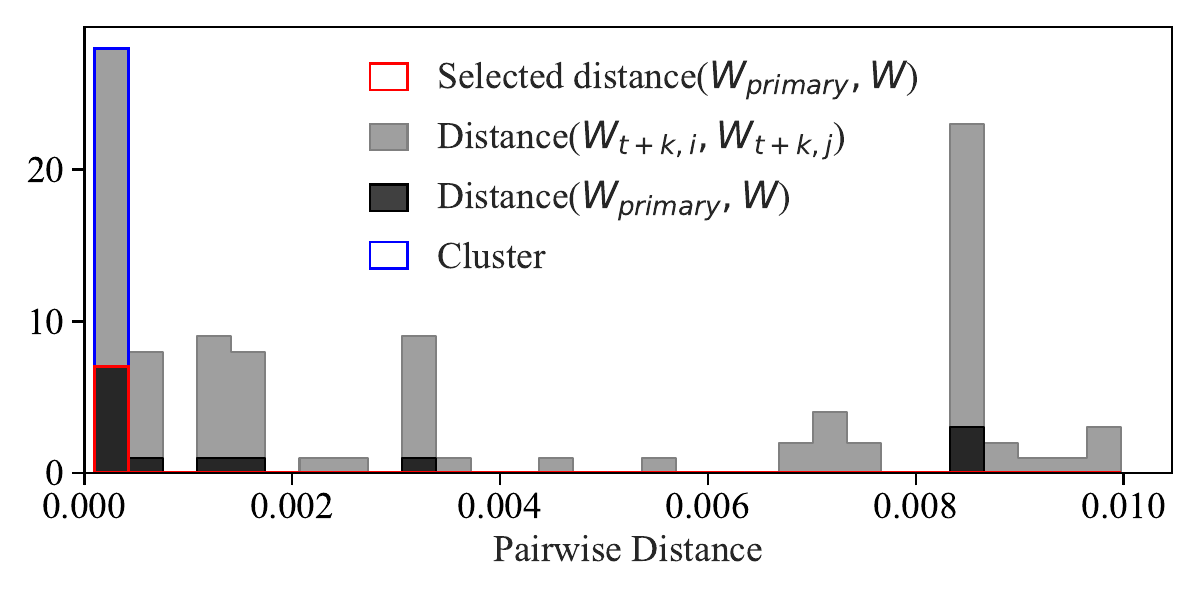}
}
\caption{
Reproduction of \Cref{fig:main} on VGG-19 trained on the ImageNet dataset. The results are consistent. Note that the experiments here are performed by continuing training on a pretrained VGG-19. We consider the pretrained model as the checkpoint selected---we perform a subrun of training from it, with $k=5000$ (approximately 1 epoch of training on the ImageNet dataset). We modified the backdoor attacks to adapt to ImageNet, as most of them were not tested on the complete ImageNet dataset, and made sure the attack success rates are significantly above random guessing. However, we were not able to do this for the BppAttack as its official implementation raises out-of-memory errors when adapting to ImageNet.
}
\label{fig:imagenet}
\end{figure}

\begin{figure}[h!]
\centering
\subfloat[Pairwise Distance
]
{
\includegraphics[width=0.95\linewidth]
{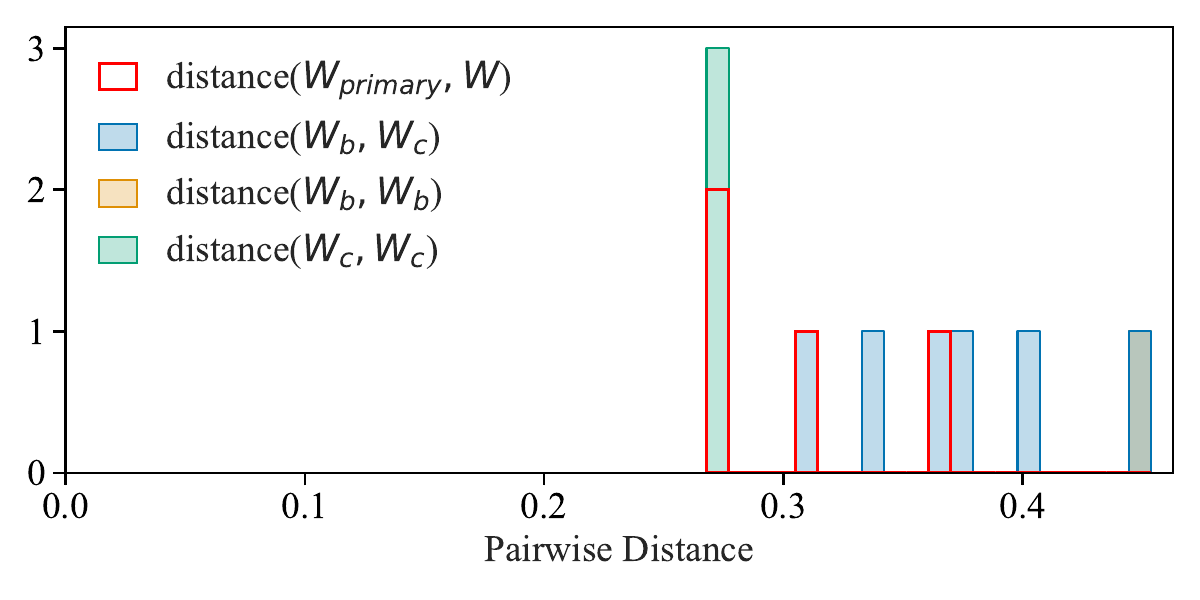}
}

\subfloat[Client's View
]
{
\includegraphics[width=0.95\linewidth]
{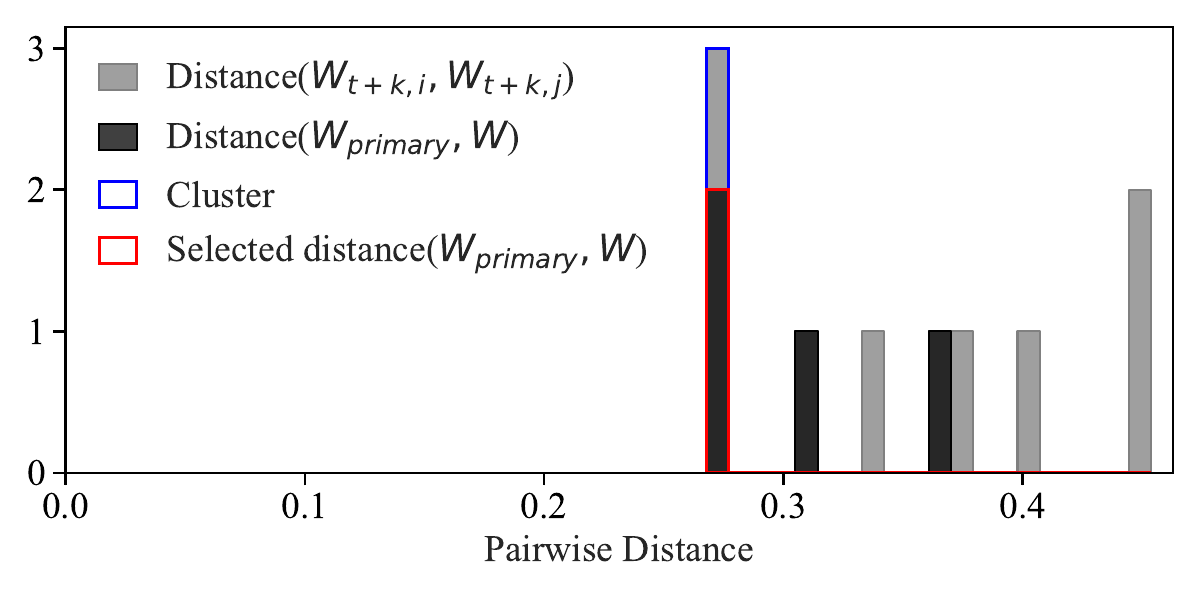}
}
\caption{
Reproduction of \Cref{fig:main} on BERT models trained on the AG news dataset. The results are consistent.
}
\label{fig:bert}
\end{figure}

\begin{figure}[h!]
\centering
\subfloat[Benign Primary Server
]
{
\includegraphics[width=\linewidth]{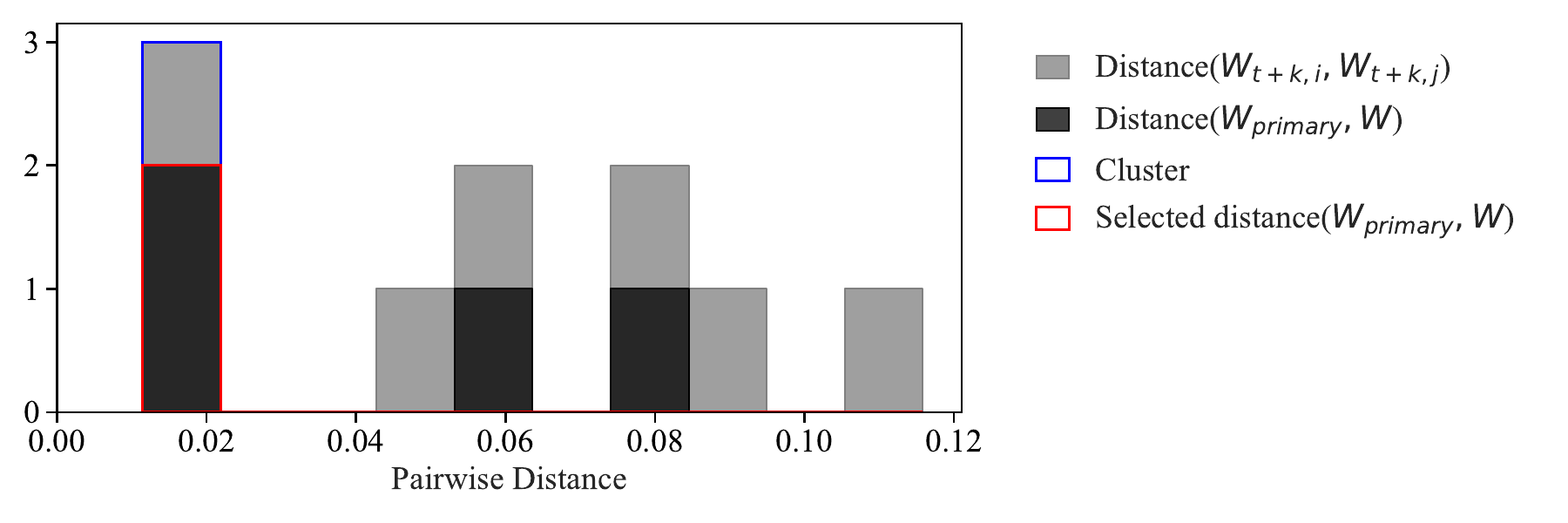}
}

\subfloat[Malicious Primary Server
]
{
\includegraphics[width=\linewidth]{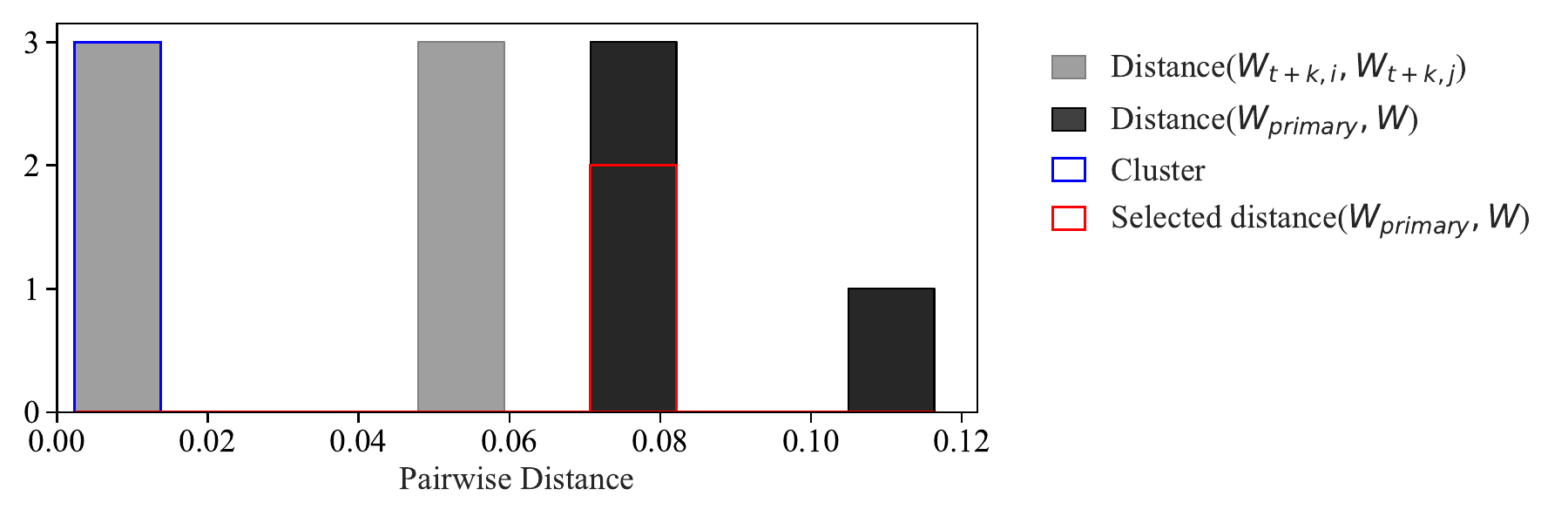}
}
\caption{Reproduction of Figure~\ref{fig:main} when there are only 5 training servers. Figure is from the client's point of view, and shows both a benign and malicious primary server. The results are consistent--distances from benign servers fall into the cluster of distance($W_c, W_c$), and vice versa for malicious servers.}
\label{fig:5servers}
\end{figure}

\clearpage

\end{document}